\documentclass{article}

\usepackage{arxiv}

\usepackage[utf8]{inputenc} 
\usepackage[T1]{fontenc}    
\usepackage{hyperref}       
\usepackage{url}            
\usepackage{booktabs}       
\usepackage{amsfonts}       
\usepackage{nicefrac}       
\usepackage{microtype}      
\usepackage{lipsum}
\usepackage{amsmath}
\usepackage{pgf,tikz,pgfplots}

\usepackage{color}
\usepackage[utf8]{inputenc}
\usepackage[T1]{fontenc}
\usepackage[caption=false]{subfig}
\usepackage{notoccite}
\usepackage{longtable}
\usepackage{pgf,tikz,pgfplots}
\usepackage{mathrsfs}
\usetikzlibrary{arrows}
\usepackage{enumitem}
\usepackage{graphicx}
\usepackage{dcolumn}
\usepackage{bm}

\usepackage{url}

\usepackage{color}

\usepackage[utf8]{inputenc}
\usepackage[T1]{fontenc}

\pgfplotsset{compat=1.14}

\newcommand{\mysize}{0.45}
\newcounter{undefinedreferences}
\title{Extracting chromatic properties of electron beams from spectral analysis of turn-by-turn beam position data}

\author{
 Panagiotis~Zisopoulos\\
  CERN\\
  Geneva, CH-1211 \\
  Switzerland \\
  \texttt{panagiotis.zisopoulos@cern.ch} \\
   \And
 Yannis~Papaphilippou \\
  CERN\\
  Geneva, CH-1211 \\
  Switzerland \\
}
\begin{document}
\maketitle







\date{\today}

\begin{abstract}
A method to estimate linear chromaticity, RMS energy spread, and chromatic beta-beating, directly from turn-by-turn beam position data in a circular electron accelerator, is presented. This technique is based on spectral analysis of a transversely excited beam, in the presence of finite chromaticity. Due to the turn-by-turn chromatic modulation of the beam's envelope, betatron sidebands appear around the main frequency of the Fourier spectra. By determining the amplitude of both sidebands, chromatic properties of the beam can be estimated. In this paper, analytical derivations justifying the proposed method are given, along with results from tracking simulations. To this end, results from practical applications of this technique at the KARA electron ring are demonstrated.
\end{abstract}




\section{Introduction\label{sec:Intro}}
The measurement of chromaticity~\cite{Verdier:1993iz} is an indispensable process in a circular accelerator, due to the impact it has on beam lifetime and quality. At the modern high-intensity proton and electron rings, the thresholds of several beam instabilities are controlled through the value of chromaticity. A plethora of methods for measuring chromaticity have been developed so far, each with different beam parameters as observables. Some of the existing techniques are mentioned below, however a more detailed review of the corresponding approaches can be found in~\cite{Steinhagen:1213281, Minty:2003fz}.

 One of the most simple and widely used techniques is the \emph{RF-sweep}, where the operating frequency of the radio-frequency (RF) system is changed, in order to induce a change in the energy of the beam. At the same time, betatron tune measurements~\cite{Serio:BFb0018282, Bartolini:1996gj} are performed, usually by analysing turn-by-turn (TbT) data from beam position monitors (BPMs) and employing Fourier algorithms for frequency analysis. Chromaticity can be then determined from the correlation of the measured tunes with the varying energy of the beam. This method usually requires dedicated experimental time, thus making it difficult to use during normal operation of the accelerator.
 
 The estimation of chromaticity is also possible by measuring the \emph{incoherent} power spectrum of the beam with Schottky monitors~\cite{Boussard:1993jq}. This method is particularly useful for high-intensity, high-energy proton machines, like the \emph{Large Hadron Collider} (LHC), however in the case of bunched beams, strong coherent beam modes can pollute the Schottky spectra and thus reduce the efficiency of the method~\cite{Betz:2286306}.
 
 A very interesting technique exists that allows for the measurement of the \emph{relative chromaticity} $\xi=Q'/Q$, where $Q$ is the betatron tune, by observing the phase shift of the head and the tail of a bunch~\cite{Cocq:1998bp, Fartoukh:2002ck} during one synchrotron period. This phase-shift is correlated with the value of chromaticity and it can be measured with instruments that can resolve intra-beam movements, like the \emph{Head-Tail Monitor}~\cite{Catalan-Lasheras:624191}. This method has been found to be significantly affected by the quality of the TbT signal~\cite{Ranjbar:2017msw}, where \emph{kinematic decoherence}~\cite{Meller:1987ug, Lee:1991ii} arising from the frequency distribution over the particles in the bunch, can severely diminish the signal-to-noise ratio. As a result, only a few hundreds of turns can be left for meaningful frequency analysis, whereas the synchrotron period of a proton machine can be much larger. Note that this technique cannot be easily applied in a high-energy electron ring, where the bunch lengths are in the range of picoseconds, rendering the phase differences between the head and the tail of the bunches hard to resolve.
 
 Another class of methods that can be employed are the ones that extract information from the chromatic decoherence of the beam~\cite{Siemann:1989hx}. Due to this mechanism, as the beam revolves around a circular accelerator, its' envelope is modulated in amplitude. By determining the maximum and the minimum of the coherent TbT signal i.e.~the modulation depth of the betatron oscillations, information on chromaticity and the RMS energy spread can be acquired. In fact, measuring and controlling the RMS energy spread is of paramount importance as well for the modern high-intensity electron accelerators. For example, the modern light sources are designed as to achieve very low emittances, in the regime that \emph{intra-beam scattering} becomes important~\cite{Kubo:2005rn, Bane:2002ss, Antoniou:2012zbr}. 
 At an electron ring, the RMS energy spread is typically measured by evaluating information from the synchrotron light of the electron bunches with specific instrumentation e.g. a streak camera. However, the same goal can be achieved by analysing TbT BPM data in the \emph{time domain}~\cite{Hsu:1990an,Bassi:2015rvu}, by performing multi-parametric fits, based on the existing analytical models of the beam centroid motion in the presence of chromaticity. Unfortunately, the unavoidable noise in the TbT data and small values of the chromaticity can drive large measurement errors when these methods are employed~\cite{Manukyan:2011zz}.

In this paper, by acknowledging the necessity for an operationally efficient technique, an alternative method is proposed for estimating the chromaticity or the RMS energy spread through the Fourier spectra of a \emph{transversely} excited beam. The proposed technique is based on the analytical relationships derived in~\cite{Rumolo:2002bx}, which govern the Fourier spectra of a \emph{longitudinally} excited beam, in the presence of chromaticity. The method takes advantage of the fact that chromatic decoherence modulates the envelope of the beam, with the modulation period being exactly the synchrotron period. As a result, chromatic sidebands appear around the main betatron tune in the transverse spectra of the excited beam, with an amplitude proportional to chromaticity. With the proposed method, the estimation of chromaticity is performed through the measurement of both the chromatic sidebands of the beam, and by using the following simple equation 
\begin{equation}
\label{Intro:Eq1}
Q'_z=\pm\frac{Q_{s}}{\sigma_{\delta}}\sqrt{\frac{A_{1}+A_{-1}}{A_{0}}}\,\,,
\end{equation}
where $Q'_{z}$, with $z=x, y$, is the horizontal or vertical chromaticity respectively, $Q_{s}$ is the synchrotron tune, $\sigma_{\delta}$ is the RMS energy spread, $A_{1}$ and $A_{-1}$ are the amplitudes of the first order chromatic sidebands that appear around the horizontal or vertical betatron frequency which has an an amplitude of $A_{0}$. Precise measurements of chromaticity with the aforementioned method are possible only if the Fourier amplitudes $A_{\pm}1$ are known with high certainty. Fortunately, the existence of powerful numerical tools for the Fourier decomposition of the beam's motion allow for this. The \emph{Numerical Analysis of Fundamental Frequencies} (NAFF)~\cite{Laskar:1992zz,LASKAR1993257,2003math......5364L,Papaphilippou:2014jma} algorithm has been used with success in the field of accelerator physics, for precise optics and dynamical stability measurements through frequency and amplitude analysis. The first results from the application of this method can be found in~\cite{Zisopoulos:2014hva}, where TbT RMS energy spread measurements are performed at the Swiss Light Source (SLS).

It is important to mention that a method which follows a similar approach, i.e. is based on the spectral response of the chromatic motion of the beam, has been developed in~\cite{Nakamura:871891} and employed for experimental measurements~\cite{Nakamura:1999ve, Kiselev:2007zz}. A limitation of this method is that it assumes equal amplitudes for both chromatic sidebands, and it uses only one sideband for measurements. However, this would be true only in the absence of non-linearities which can cause beating of the optics in the lattice.

 In order to by-pass this limitation, the method proposed in this paper uses both amplitudes of the chromatic sidebands as it is testified in Eq.~\eqref{Intro:Eq1}. The same methodology has been also applied to the Diamond Light Source~\cite{Rehm:2010dmn}, where a piecewise fit of the ratio of the chromatic sidebands, for various chromaticities, has been implemented and used as a reference, in order to estimate chromaticity during operation. However, in the present work, the estimation of chromaticity is model-independent, by solving analytically the equations of decoherence, and by using the amplitudes of the sidebands as an observable. An advantage of this method is that it requires the same experimental procedure as the typical betatron tune measurements in a ring. 
 
 The present paper is organised as follows: In Section~\ref{sec:Anal} the theoretical basis of this method is presented. In Section~\ref{sec:Sims}, the method is employed in tracking simulations for estimating chromaticity and chromatic beta-beating from TbT BPM data, by using the accelerator model of the KARA light source and NAFF. In Section~\ref{sec:Meas}, the proposed method is employed in experimental measurements at the KARA light source, including an application to characterize the optics response during the commissioning of the Compact Linear Collider (CLIC) Superconducting Wiggler prototype~\cite{Bernhard:IPAC2016-WEPMW002}.

 \section{Analytical Derivations\label{sec:Anal}}

\begin{figure}[!htb]
  \centering
  \includegraphics[width=\mysize\textwidth]{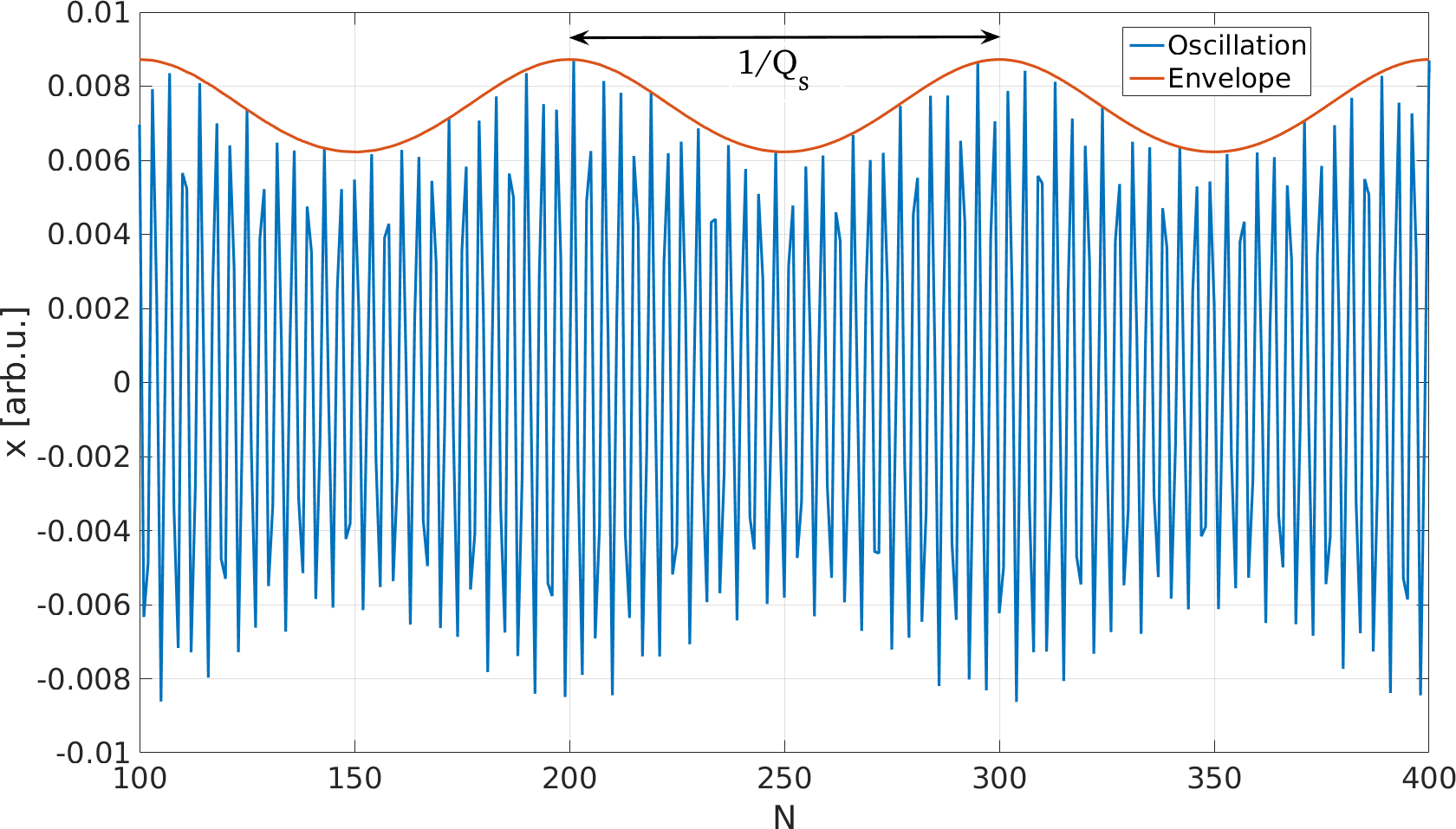}
  \caption{Example of betatron oscillations recorded at a fictitious BPM with respect to the number of turns $N$ in the presence of chromaticity. The betatron oscillations are shown in blue and the upper envelope of the signal in red. The amplitude modulation due to chromaticity has a period of $N=100$ turns. The synchrotron tune is $Q_s=0.01$. In this example, the transverse motion of the centroid is linear i.e. no amplitude detuning is present.}
  \label{sims:FIG0}
\end{figure}

Analytical formulas of the Fourier spectra of a Gaussian beam, which is longitudinally excited in the presence of chromaticity, have been derived in~\cite{Rumolo:2002bx}. These expressions take into account non-linearities, which arise from the distortion of the beam's motion due to finite dispersion at the sextupoles. From these relationships, the Fourier amplitudes $A_q$ of the synchrotron sidebands of order $q$ are given from
\begin{align}
\label{sec:Anal:eq1}
  A_q={}e^{-s^2}|\tilde{\alpha}|&\biggl|I_q(s^2-isk)+\frac{\Delta\beta_{z}}{4i\beta_{z}}\sigma_{\delta}(k+is)\\ \nonumber
  &\bigl[I_{q-1}(s^2-isk)-I_{q+1}(s^2-isk)\bigr]\biggr|\,\,,
\end{align}
where $s=Q'_z\sigma_{\delta}/Q_s$ with $Q'_z$ the chromaticity for the $z=x,~y$ planes, $\sigma_{\delta}$ the RMS energy spread of the bunch and $Q_s$ the synchrotron tune, $\tilde{\alpha}$ is the initial transverse amplitude of the kicked beam, $I_q(x)$ is the $q$-order modified Bessel function of the first kind with argument $x$, $k$ is the longitudinal kick in units of the RMS bunch length $\sigma_t$ of the beam, and $\frac{\Delta\beta_{z}}{\beta_{z}}$ is the chromatic beta beating which characterizes the dependence of the beta function $\beta_z$ on the energy deviation of the beam $\delta$. 
The appearance of the $I_q(s^2-isk)$ term is a signature of amplitude modulation, where the argument $s^2-isk$ represents the \emph{modulation index}~\cite{Oppenheim:107704}. The same modulation occurs in the case of a pure transverse excitation. A numerical example is shown in Fig.~\ref{sims:FIG0}, where TbT pseudo-data are produced, based on the kinematic decoherence relationships in~\cite{Rumolo:2002bx}. The envelope (red curve) of the oscillation (blue curve) exhibits a modulation period of $N_{s}=100$ turns i.e. the inverse of the synchrotron tune $Q_{s}=0.01$. Every $N_{s}/2$ turns the envelope passes a trough, thus exhibiting a minimum projection of the synchrotron oscillations in the transverse betatron motion. The carrier frequency is the betatron frequency, which is excited with a transverse impulse at $N=0$ turns.

For the application presented in this paper, the longitudinal excitation is dropped in favor of a transverse one. The motivation behind this choice is to develop a method for chromaticity or RMS energy spread measurements, based on the same procedure of transverse betatron tune measurements. In order to do so, the parameter of the longitudinal excitation $k$, is set to $0$ and Eq.~\eqref{sec:Anal:eq1} becomes
\begin{equation}
  A_q=e^{-s^2}|\tilde{\alpha}|\biggl|I_q(s^2)+s\frac{\Delta\beta_{z}}{4\beta_{z}}\sigma_{\delta}\bigl[I_{q-1}(s^2)-I_{q+1}(s^2)\bigr]\biggr|\,\,,
  \label{sec:Anal:eq2}
\end{equation}
where now the Fourier amplitudes of the beam $A_q$, are determined from the transverse impulse given to the beam. As a matter of fact, one of the assumptions of the current study is that the transverse kick given to the beam is adequately small, such as not to induce rapid decrease of the beam's centroid oscillation due to decoherence, as a result of tune-shift with amplitude.

The previous expression can be written in a simpler form by using the following relationships~\cite{Abramowitz}:

\begin{align}
&I_{q-1}(x)-I_{q+1}(x)=\frac{2q}{x}I_q(x)\label{sec:Anal:eq3} \\
&I_q(x)=I_{-q}(x) \label{sec:Anal:eq4} \\ 
&I_q(x)\approx\frac{1}{\Gamma(q+1)}\biggl(\frac{x}{2}\biggr)^q,\quad 0<|x|\ll \sqrt{q+1}\label{sec:Anal:eq5}\,\,.
\end{align}

Indeed, by using Eq.~\eqref{sec:Anal:eq3} in Eq.~\eqref{sec:Anal:eq2} and computing the absolute value of the result, the expression 

\begin{equation}
  A_q=|\tilde{a}|e^{-s^2}I_q(s^2)\biggl(1+q\frac{Q_s}{Q'_z}\frac{\Delta\beta_{z}}{2\beta_{z}}\biggr)\,\,,
  \label{sec:Anal:eq6}
\end{equation}
is obtained. 

It is evident from Eq.~\eqref{sec:Anal:eq6} that chromatic beta-beating becomes significant for higher-order modes, i.e. it is linear with $q$. In addition, it is clear that by using only one synchrotron sideband for the estimation of the chromatic properties induces an uncertainty due to the beta-beating. However, by forming the sum of both symmetric sidebands $A_{q}+A_{-q}$ with the help of Eq.~\eqref{sec:Anal:eq4}, and normalising it with the main betatron amplitude $A_0$ for $q=0$, results in the elimination of the perturbation term $\frac{\Delta\beta}{2\beta}$. Moreover, by normalising the difference $\pm|A_q-A_{-q}|$ by $A_{q}+A_{-q}$, an analytical relationship for the estimation of the chromatic beta-beating $\frac{\Delta\beta}{\beta}$ is obtained. The aforementioned operations result in the following equations:

\begin{align}
  \frac{A_q+A_{-q}}{A_0}&=\frac{2I_q(s^2)}{I_0(s^2)}\label{sec:Anal:eq7} \\
  \biggl|\frac{A_q-A_{-q}}{A_q+A_{-q}}\biggr|&=\biggl|\pm\frac{q\,Q_s}{2\,Q'_z}\frac{\Delta\beta_z}{\beta_{z}}\biggr|\label{sec:Anal:eq8}\,\,.
\end{align}

Apart from Eq.~\eqref{sec:Anal:eq8} which can be directly used to estimate the chromatic beta-beating $\frac{\Delta\beta_z}{\beta_{z}}$, the linear chromaticity $Q'_{z}$ can be inferred by solving numerically Eq.~\eqref{sec:Anal:eq7} in order to estimate the $s=\frac{Q'_{z}\sigma_{\delta}}{Q_{s}}$ parameter. The synchrotron tune $Q_{s}$ is usually known with fair accuracy from the properties of the RF system, but it can also be inferred from the frequency offset of the chromatic sidebands, with respect to the main frequency line. Then, by knowing the RMS momentum spread $\sigma_{\delta}$, one can estimate the chromaticity $Q'_{z}$ of the machine or \textit{vice-versa}. 

 Although numerical solutions of Eq.~\eqref{sec:Anal:eq7} can be trivially produced with the help of modern numerical libraries~\cite{2020SciPy-NMeth}, simple analytical relationships that determine the chromaticity or the RMS energy spread can be acquired by introducing the approximation in Eq.~\eqref{sec:Anal:eq5}. Assuming that $s^{2}\ll\sqrt{q+1}$, for the $q$ order chromatic sideband under consideration, the analytical form of Eq.~\eqref{sec:Anal:eq7} becomes 

\begin{align}
\frac{A_q+A_{-q}}{A_0}&\approx \frac{2}{\Gamma(q+1)} \biggl(\frac{s^{2}}{2}\biggr)^{q} \nonumber\\
&\approx\frac{2^{{1-q}}}{\Gamma(q+1)}s^{2q}\label{sec:Anal:eq9}\,\,,~\text{subject to}~s^{2}\ll\sqrt{q+1}\,\,.
\end{align}

Solving Eq.~\eqref{sec:Anal:eq9} with respect to $s$, and furthermore with respect to the chromaticity $Q'_{z}$, yields

\begin{equation}
Q'_{z}\approx \pm \frac{Q_{s}}{\sigma_{\delta}}\,2^{\frac{q-1}{2q}}\,\sqrt[2q]{\Gamma(q+1)\biggl[\frac{A_{q}+A_{-q}}{A_{0}}\biggr]}
\label{sec:Anal:eq10}
\end{equation}

The previous expression allows for the estimation of chromaticity $Q'_z$ or the RMS energy spread $\sigma_{\delta}$, based on the amplitudes of the chromatic sidebands $A_{q}$ and $A_{-q}$, for $q\geq1$. Since in a real Fourier spectrum of TbT data, assuming that the second order chromaticity $Q''_{z}$ is not large, the beam is longitudinally matched to the RF bucket, and that collective effects are not important, the first order chromatic sidebands $A_1$ and $A_{-1}$ are most easily resolved. Thus, by setting $q=1$ in Eq.~\eqref{sec:Anal:eq10} and in Eq.~\eqref{sec:Anal:eq8}, the following expressions are obtained:

\begin{align}
\label{sec:Anal:eq11}
Q'_{z} &\approx \pm \frac{Q_{s}}{\sigma_{\delta}}\sqrt{\frac{A_{1}+A_{-1}}{A_{0}}}  \\ 
\label{sec:Anal:eq12}
\frac{\Delta\beta_z}{\beta_{z}}&=\pm \frac{2\,Q'_{z}}{Q_{s}}\biggl|\frac{A_1-A_{-1}}{A_1+A_{-1}}\biggr|\,\,,
\end{align}
where according to the constraint in Eq.~\eqref{sec:Anal:eq9}, Eq.~\eqref{sec:Anal:eq11} is valid if
\[
 Q_z'\ll2^{\frac{1}{4}}\,\frac{Q_s}{\sigma_{\delta}}~.
 \] 
 For the case of an electron ring, typical values of the synchrotron tune ($Q_s\approx10^{-2}$) and the RMS energy spread ($\sigma_{\delta}\approx10^{-3}$) give an approximate upper bound in the values of chromaticities that can be inferred ($Q_z'\ll10$) with the previous approximation, Eq.~\eqref{sec:Anal:eq11}.

 The relationships in Eq.~\eqref{sec:Anal:eq11} and Eq.~\eqref{sec:Anal:eq12} are independent from the calibration of the BPMs due to normalization to the main amplitude $A_0$, for estimating the chromaticity or the RMS energy spread of the beam. As a by-product of the method, a relationship which determines the chromatic beta-beating at the position of the BPMs is recovered. The previous expressions do not differentiate between positive and negative solutions, since they depend on the Fourier amplitudes of the beams. Thus, the proposed method cannot recover the sign of the chromaticity $Q'_{z}$ and of the chromatic beta beating $\frac{\Delta\beta_z}{\beta_{z}}$. Analytical expressions for the measurement errors of chromaticity and chromatic beta-beating through Eq.~\eqref{sec:Anal:eq11} and Eq.~\eqref{sec:Anal:eq12}, can be found in Sec.~\ref{sec:app:error} of the Appendix.

\section{Simulations\label{sec:Sims}}

\begin{figure}[!htb]
  \centering
  \includegraphics[width=\mysize\textwidth]{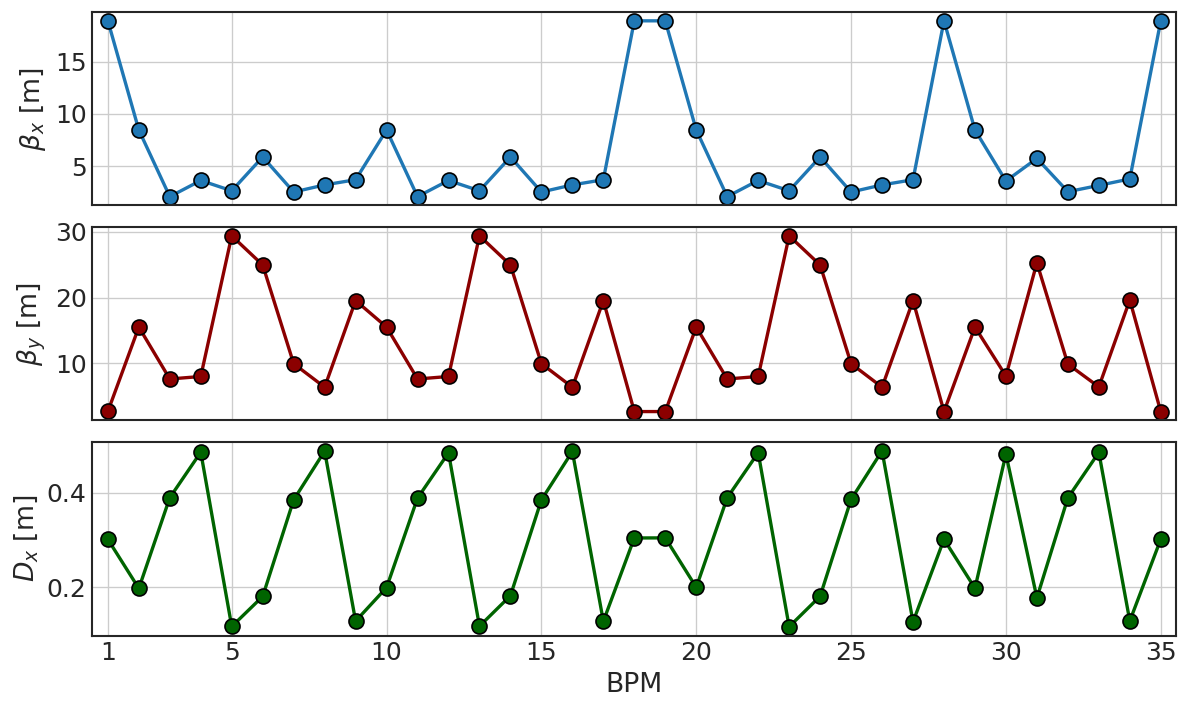}
  \caption{The optics for the accelerator model of the KARA light source, with respect to the position of the $M=35$ BPMs of the KARA light source. Top and middle plots show the horizontal $\beta_x$ and vertical $\beta_y$ functions, and the bottom plot depicts the horizontal dispersion $D_x$.}
  \label{sims:FIG1}
\end{figure}

Tracking simulations with the model of the KARA electron accelerator are undertaken using \emph{MADX-PTC}~\cite{Schmidt:2002vp}, in order to numerically investigate the proposed method and to identify dependence on the initial excitation of the beam. The fundamental parameters of the simulations can be found in Table.~\ref{table1}. The particles are tracked around the KARA lattice for different cases of initial excitation, and for $N=1200$ turns. The lattice of the KARA accelerator consists of Double Bend Achromat (DBA) cells and exhibits a four-fold symmetry. The number of BPMs is $M=35$ and the optics at the position of the BPMs are shown in Fig.~\ref{sims:FIG1}, where the maximum horizontal beta function (top) is around $\beta_{x}=19$~m at the position of the arcs of the ring, the maximum vertical beta function is approximately $\beta_{y}=30$~m and the horizontal dispersion (bottom) exhibits an average value of around $\langle D_{x}\rangle=0.25$~m. The simulations employ an error-free lattice, with no radiation effects taken into consideration, which result in a vanishing vertical dispersion $D_y$ around the accelerator.


\begin{table}[htb!] 
  \small
  \centering
  \caption{Parameters of the tracking simulations with the KARA model. All the position dependent parameters are measured at the injection point of the ring.}
  \begin{tabular}{l |l}
      \toprule
\textbf{Parameter} & \ \ \ \textbf{Value} \\
\toprule
       Energy           &  \ \ \ 2.50~[GeV]        \\ 
       Circumference    & \ \ \ 110.40~[m] \\
       Tune  Q$_x$, Q$_y$, $Q_{s}$    & \ \ \ 6.79, \ 2.70, \ 0.016~[2$\pi$]~      \\
     Chromaticity $Q'_x$, $Q'_y$      & \ \ \  2.57, \ 6.90~[2$\pi$]      \\
     RMS beam size $\sigma_x$, $\sigma_y$ at injection    & \ \ \  0.94, 0.080~[mm]  \\
     Beta function $\beta_x$, $\beta_y$ at injection & \ \ \ 18.89, 1.67~[m] \\
     RMS bunch length $\sigma_{z}$  & \ \ \ 10.30~[mm]   \\
     RMS energy spread $\sigma_{\delta}$ &\ \ \ 9.80~[$10^{-4}$] \\
     Initial excitation at injection  & \ \ \ $1-7$ [$\sigma_x$], $20-80$ [$\sigma_y$] \\
     Distribution & \ \  \ Gaussian \\

 \toprule
  \end{tabular}
  \label{table1}
\end{table}

\begin{figure}[!htb]
\centering
\subfloat[Horizontal TbT data for excitations of $1~\sigma_{x}$ (top left), $2~\sigma_{x}$ (top right), $3~\sigma_{x}$ (bottom left) and $4~\sigma_{x}$ (bottom right).\label{sims:FIG2}]{\includegraphics[width=\mysize\textwidth]{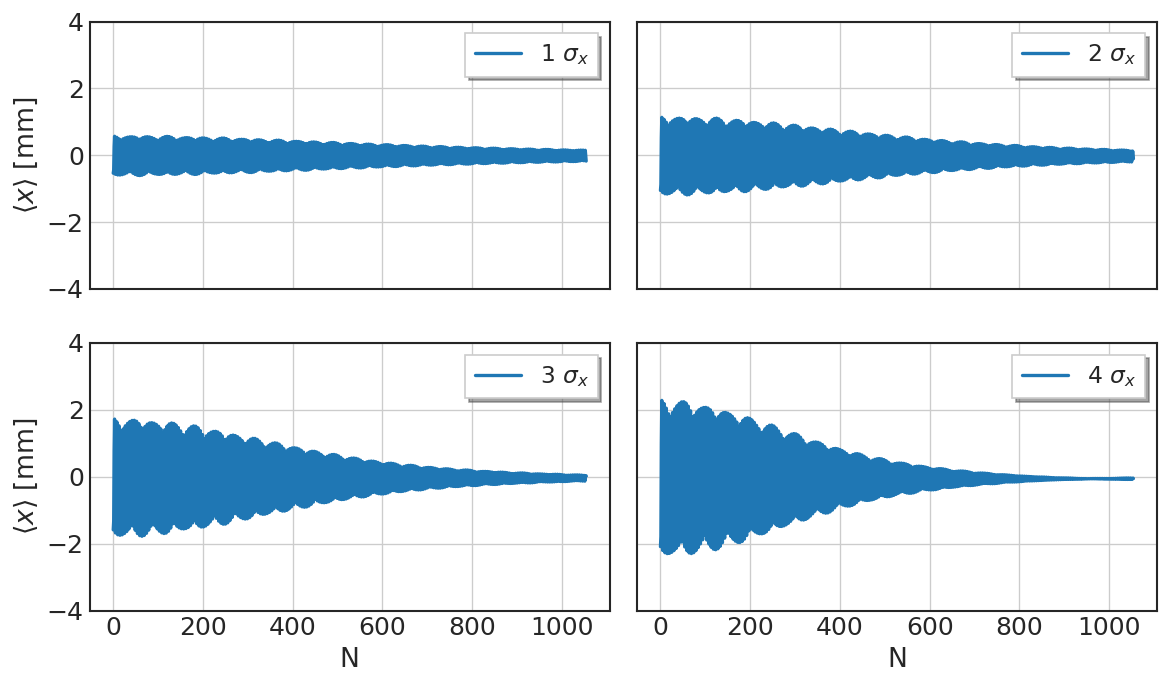}} \\
\subfloat[Vertical TbT data for excitations of $20~\sigma_{y}$ (top left), $30~\sigma_{y}$ (top right), $40~\sigma_{y}$ (bottom left) and $50~\sigma_{y}$ (bottom right).\label{sims:FIG3}]{\includegraphics[width=\mysize\textwidth]{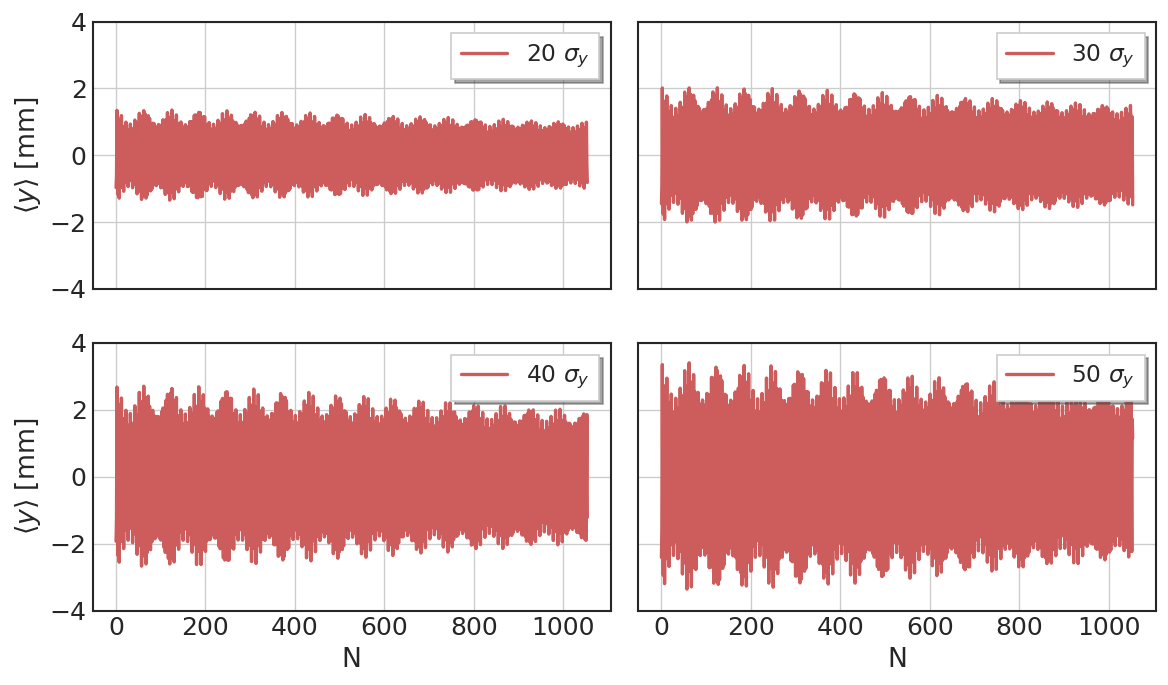}}
\caption{The simulated TbT tracking data of the beam's centroid for the KARA accelerator model, with respect to the number of turns $N$. The initial amplitude of the beam, in terms of the transverse beam sizes $\sigma_x$, $\sigma_y$, is indicated in the legend.}
\end{figure}

The TbT evolution of the centroid of the beam, is inferred from the arithmetic mean of all the particles for each turn $N$. The beam is transversely excited in a range of initial amplitudes, which are referred in this paper in units of the RMS transverse beam size $\sigma_z$, where $z=x,~y$ for the horizontal and vertical planes respectively. In Fig.~\ref{sims:FIG2}, the horizontal TbT are shown for initial excitations of 1~$\sigma_{x}$ to 4~$\sigma_{x}$, where for that particular position in the ring, the initial amplitude ranges from $0.5$~mm to $2$~mm. The TbT simulated data for the vertical plane are presented in Fig.~\ref{sims:FIG3}, after initial excitations of 20 to 50~$\sigma_{y}$ and initial amplitudes similar to the horizontal case in Fig.~\ref{sims:FIG2}. For both planes, the observed damping of the beam's envelope is caused by the tune-shift with amplitude induced by the sextupoles in the KARA lattice, whereas the fast TbT beating of the beam's centroid is caused by the chromatic oscillations. The trend of the centroid's evolution in the vertical plane exhibits a smaller non-linear detuning and larger chromatic oscillations, compared to the horizontal plane, due to the higher vertical chromaticity. 

\subsection{Envelope evolution}

\begin{figure}[!htb]
\centering
\subfloat[\label{sims:FIG4} Horizontal normalized envelopes of the centroid for the case of 1~$\sigma_{x}$ (blue line), 2~$\sigma_{x}$ (orange line), 3~$\sigma_{x}$ (green line), and 4~$\sigma_{x}$ (red line).]{\includegraphics[width=\mysize\textwidth]{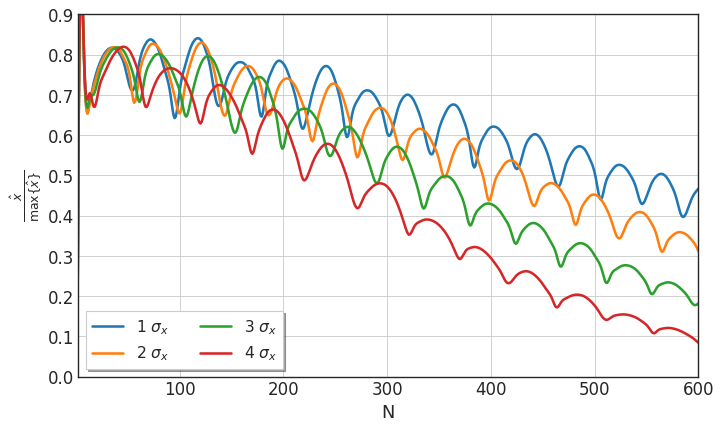}} \\
\subfloat[\label{sims:FIG5} Vertical normalized envelopes of the centroid for the case of 20~$\sigma_{y}$ (blue line), 30~$\sigma_{y}$ (orange line), 40~$\sigma_{y}$ (green line), and 50~$\sigma_{y}$ (red line).]{\includegraphics[width=\mysize\textwidth]{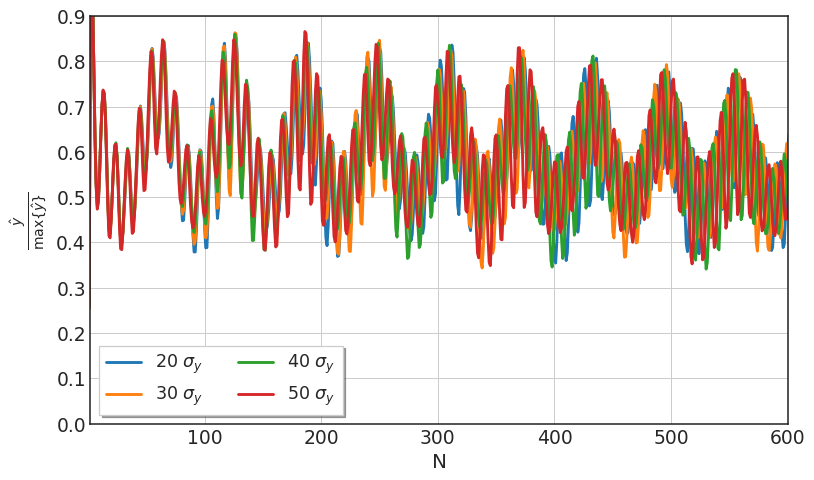}}
\caption{The upper envelopes for the horizontal (a) and vertical (b) oscillations of the centroid of the beam for the KARA model, with respect to the number of turns $N$. The envelopes are normalised to the oscillation value at the first turn $N=1$. The initial excitation for each simulation is shown in the legend.}
\end{figure}

As it has been discussed already, chromatic decoherence results in the periodic modulation of the beam's envelope with synchrotron period $\tau_{s}=\frac{1}{Q_{s}}$. For the present simulations, the TbT evolution of the horizontal envelopes, normalised to their maximum value at $N=1$ turns, is shown in Fig.~\ref{sims:FIG4} for a range of initial excitations. The synchrotron period is about $\tau_s=63$ turns and during the first synchrotron period, the centroid of the beam oscillates with the same phase, regardless of the initial excitation. However, during and after the second synchrotron period, a distinct amplitude-dependent damping of the synchrotron oscillations appears. This behaviour can be attributed to the non-linear detuning of the centroid, which becomes dominant after around $N=300$ turns, i.e. almost $5$ synchrotron periods $\tau_s$. In addition, the phases of the oscillations for all cases appear to become different after the first two synchrotron periods $\tau_s$. This indicates that the synchrotron oscillations exhibit a frequency shift, which depends on the initial excitation amplitude. For the $4~\sigma_{x}$ case and for that particular BPM, the value of the envelope is at 50~\% of the initial amplitude at around $N=300$ turns already, indicating a relatively small decoherence time. At the same number of turns, $N=300$, the amplitude of the horizontal oscillations for the $1~\sigma_{x}$ excitation is at about 70~\% of the initial amplitude.

The vertical envelopes are shown in Fig.~\ref{sims:FIG5}. The amplitude modulation is visible for every $N=\tau_s$ turns, with the modulation occurring around the high-frequency component i.e. the betatron oscillations. Note that the chromatic oscillations are visible also in the vertical plane, even if there is no vertical dispersion. Due to the small vertical emittance, the TbT evolution of the vertical envelopes exhibit more coherent behaviour with respect to the horizontal plane, and the dephasing of the oscillations due to decoherence begins at around five synchrotron periods. As a result, estimating the vertical chromaticity with the proposed method, is expected to be less dependent on the initial excitation. In addition, the damping of the chromatic oscillations appears to be very slow, which results in a wider range of number of turns $N$ for the survival of the synchrotron oscillations. 

An immediate conclusion of the analysis the TbT evolution of the envelopes, is that the applicability of Eq.~\eqref{sec:Anal:eq11} and Eq.~\eqref{sec:Anal:eq12} is expected to depend on the number of turns $N$, and the initial transverse excitation of the beam. For this reason, the number of turns $N$ should be restricted to as small values as possible. At the same time, that particular number of turns $N$ should allow for enough resolution for the estimation of the Fourier spectra.

\subsection{Fourier spectra}

\begin{figure}[!htb]
\centering
\subfloat[\label{sims:FIG6}Horizontal Fourier spectra, where the $2~\sigma_x$ case is shown in blue and the $4~\sigma_x$ case in orange.]{\includegraphics[width=\mysize\textwidth]{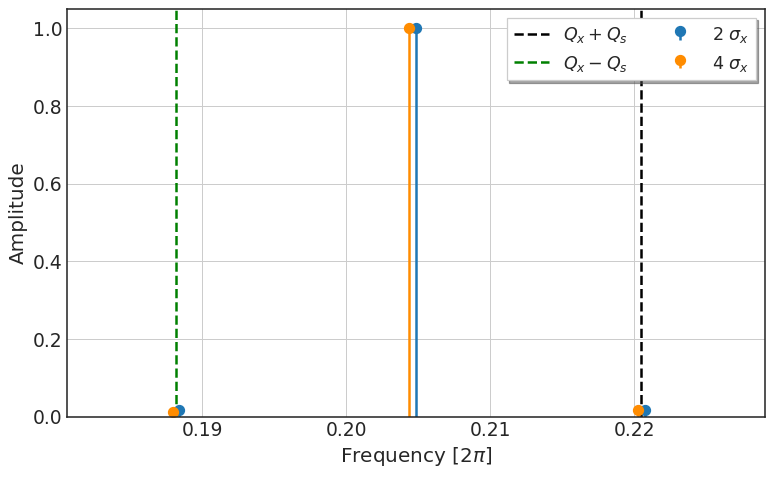}} \\
\subfloat[\label{sims:FIG7}Vertical Fourier spectra, the $30~\sigma_y$ case is shown in blue and the $50~\sigma_y$ case in orange.]{\includegraphics[width=\mysize\textwidth]{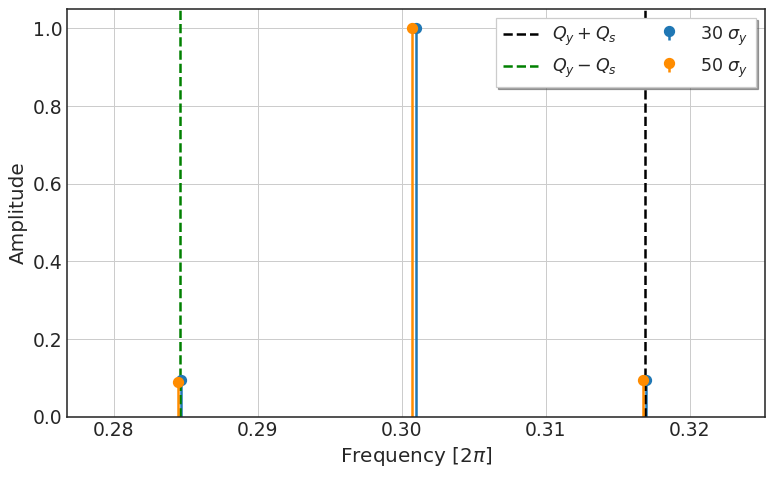}}
\caption{The frequency spectra of the beam's centroid for the horizontal~(a) and vertical~(b) planes, normalised to the amplitude of the main peak $A_0$. The chromatic sidebands $A_1$ and $A_{-1}$ are observed at a distance equal to the synchrotron tune $Q_s$, marked with dashed lines (green for $A_{-1}$ and black for $A_1$). The initial excitation amplitude of the beam, in terms of the transverse beam sizes $\sigma_x$ and $\sigma_y$, is indicated in the legend.}
\end{figure}

The Fourier spectra of the centroid of the beam are determined with \emph{PyNAFF}~\cite{PyNAFF}, the adaptation of the \emph{NAFF} \cite{Laskar:2003zz} algorithm in \emph{Python} programming language, which offers an increased precision in the estimation of the spectral components of a quasi-periodic signal. The appearance of the chromatic sidebands around the main betatron frequency line is shown in Fig.~\ref{sims:FIG6} for the horizontal plane, and in Fig.~\ref{sims:FIG7} for the vertical plane. The amplitudes are normalised to the maximum of the main Fourier amplitude $A_0$ for both cases of initial excitation, and the measurements are performed with $N=300$ turns or almost five synchrotron periods $\tau_s$.

Concerning the horizontal case, a slight increase of the horizontal tune $Q_x$ is observed for the $4~\sigma_x$ case, with respect to the $2~\sigma_x$ case, due to tune-shift with amplitude. Note that the tune-shift appears as a decrease of the horizontal betatron tune $Q_x$ in the Fourier spectra, since the fractional part of the tune $Q_x$ is larger than 0.5. 

For the vertical case, the chromatic sidebands are found to be one order of magnitude larger in amplitude, than the horizontal sidebands, due to the difference in the size of the two transverse chromaticities. Furthermore, there is a negligible shift of the frequencies for the $50~\sigma_y$ case. The frequency spectra of all the simulated cases can be similarly produced, and by using the chromatic sidebands chromaticity can be inferred by numerically solving Eq.~$\eqref{sec:Anal:eq7}$ or by directly using the approximation in Eq.~\eqref{sec:Anal:eq11}. 

\subsubsection{Synchrotron tune\label{subsub:qs}}
\begin{figure}[!htb]
  \centering
  \includegraphics[width=\mysize\textwidth]{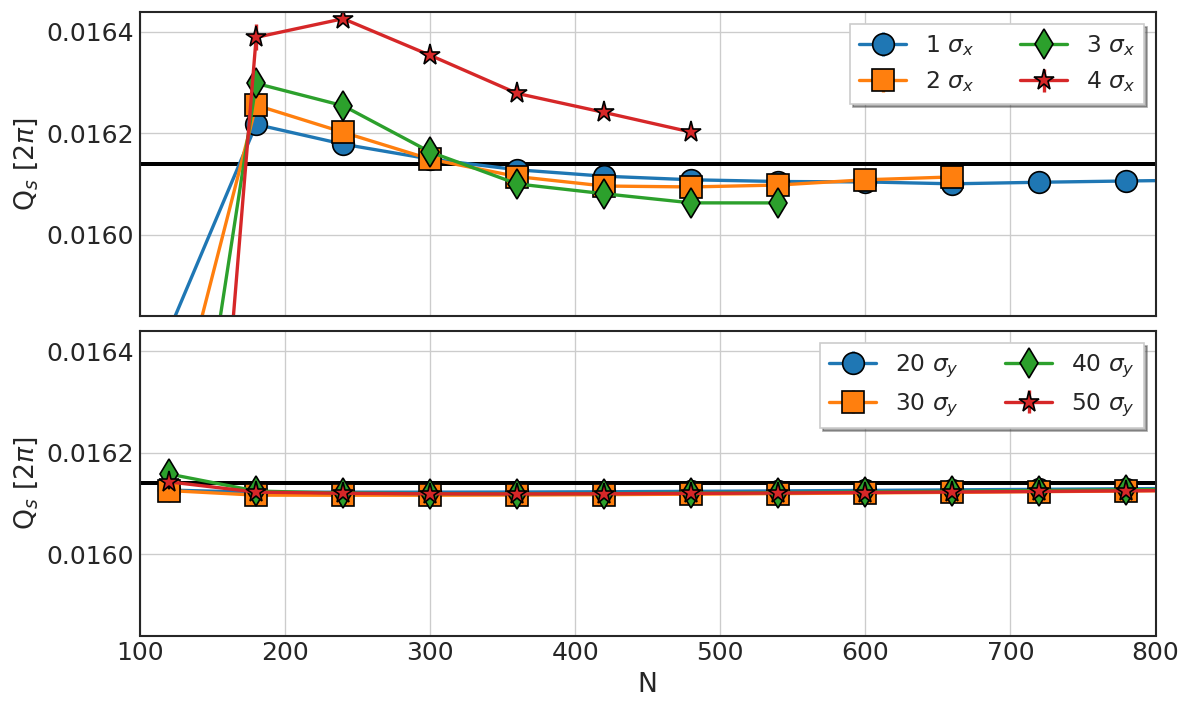}
  \caption{The synchrotron tune $Q_s$ measured from the Fourier spectra for each case of initial excitation, and averaged over all the BPMs of the KARA model. The standard deviation $\sigma_{Q_s}$ of the tunes for all cases is at the order of $\sigma_{Q_s}=10^{-7}$. The tunes are shown with respect to the number of turns $N$ for the horizontal (top) and vertical (bottom) planes. The legend indicates the initial excitation amplitude, while the black solid line shows the theoretical synchrotron tune $Q_s$, which is obtained in \emph{MAD-X}.}
  \label{sims:FIG8}
\end{figure}

 The synchrotron tune $Q_s$, is usually well known from the parameters of the RF system, however it can be also estimated from the distance between the main betatron frequency line $A_0$ and the first order chromatic sidebands $A_1$ and $A_{-1}$. Although for the present study the value of the synchrotron tune $Q_s$ is obtained from \emph{MAD-X}, it can be also estimated from the response of the transverse Fourier spectra of the beam. Such an analysis can provide useful insights on the behaviour of the TbT data. The estimation of the average synchrotron tune $Q_s$, where the average is performed on the $M=35$ BPMs of the KARA model, is graphically presented in Fig.~\ref{sims:FIG8} against the value from \emph{MAD-X} (solid black line) for the horizontal (top) and vertical (bottom) planes, and for various initial excitations.

For the horizontal plane, it is evident that amplitude dependent effects can impact the survival of the chromatic sidebands. For example, while for the $1~\sigma_x$ case the synchrotron tune can be recovered for up to $N=800$ turns, the 4~$\sigma_x$ case exhibits synchrotron sidebands up to $N=500$ turns. In addition, a longitudinal tune-shift with amplitude is evident during the first six synchrotron periods i.e. up to $N=300$ turns, which is highlighted at $N=180$ turns, and converges to a constant value after the \emph{characteristic decoherence time}~\cite{Meller:1987ug} of each initial excitation. The nature of this effect stems from the injection of the beam in the RF bucket. A larger transverse excitation results in a beam which covers more space in the longitudinal phase space, resulting in larger longitudinal emittance and synchrotron tune $Q_s$. 

The synchrotron tune inferred from the vertical TbT data exhibits no amplitude-dependent effects, and the extracted value remains constant with respect to the number of turns, and close to the expected synchrotron tune.

The TbT dependency of the synchrotron tune $Q_s$ on the initial excitation can introduce systematic errors in the estimation of chromaticity with Eq.~\eqref{sec:Anal:eq11}, which is more pronounced for stronger initial excitations. On the other hand, larger initial excitations might be favorable for resolving the chromatic sidebands. In order to correct for this source of systematic error in the synchrotron tune measurement, analytical models can be employed which describe the relationship of the synchrotron tune and amplitude. However, if this error is not large, the value of the theoretical synchrotron tune, or the value estimated with the parameters of the RF system, could be used for chromaticity measurements.

\subsubsection{Chromatic sidebands from TbT data\label{sims:sec:sb}}

In both Eq.~\eqref{sec:Anal:eq7} and Eq.~\eqref{sec:Anal:eq11}, chromaticity is obtained from the ratio of the sum of the chromatic sidebands $A_{\pm1}$, to the amplitude of the main frequency $A_0$. Due to the effect of decoherence, the amplitudes of the spectral lines decrease with respect to the number of turns $N$. The rate at which the spectral lines decrease is similar for both chromatic sidebands of order $q=\pm1$. However, due to non-linearities, it is not guaranteed that this rate is similar to the TbT decrease of the main betatron amplitudes of order $q=0$. As a consequence, different initial excitations on the beam might drive different TbT responses of the chromatic ratios. In addition, altering the initial excitation of the beam changes the decoherence profile and it thus creates unequal chromatic ratios for each excitation and at the same number of turns $N$. As a result, both of these mechanisms can reduce accuracy by introducing systematic errors in the estimation of any beam parameter from the spectral lines of the beam. 

At this point, it is worth mentioning that combined decoherence models, which describe the motion of the centroid of the beam in the simultaneous presence of finite chromaticity and tune-shift with amplitude, do exist and could be "fed-back" to the data, in order to correct for the aforementioned turn-by-turn error. This would require a proper tuning of the model and a good knowledge of the optics, as to fit the experimental data as good as possible.

For the sake of estimating the aforementioned effects, an analysis of the spectral lines of the beam's centroid for various initial excitations is employed in these simulations. First the quantities
  
\begin{align}
\label{sec:anal:eq13}
R_x&=\frac{A_{1x}+A_{-1x}}{A_{0x}} \\
R_y&=\frac{A_{1y}+A_{-1y}}{A_{0y}}
\end{align}
are defined, where $A_{1x}$, $A_{-1x}$, $A_{1y}$, $A_{-1y}$ are the first order chromatic sidebands of the main betatron amplitudes $A_{0x}$, $A_{0y}$ of the horizontal and vertical planes respectively. 

\begin{figure}[!htb]
\centering
\subfloat[\label{sims:FIG9}The evolution of the average horizontal chromatic ratio $\langle R_x \rangle$ for excitations of 1~$\sigma_x$ in blue, 2~$\sigma_x$ in orange, 3~$\sigma_x$ in green, and 4~$\sigma_x$ in red. The theoretical value is shown in black, while the uncertainty around this value shown in light green.]{\includegraphics[width=\mysize\textwidth]{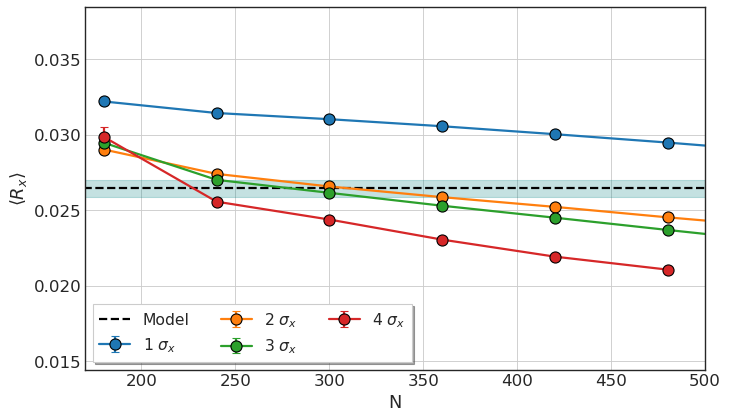}} \\
\subfloat[\label{sims:FIG10}The evolution of the average vertical chromatic ratio $\langle R_y \rangle$ for excitations of 20$~\sigma_y$ in blue, 30~$\sigma_y$ in orange, 40~$\sigma_y$ in green, and 50~$\sigma_y$ in red. The theoretical value is shown in black, while the uncertainty around this value due to statistical errors, is shown in light green.]{\includegraphics[width=\mysize\textwidth]{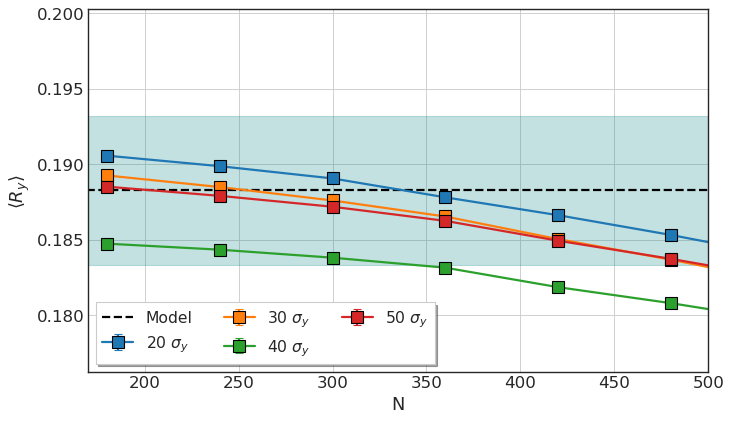}}

\caption{The TbT evolution of the chromatic ratios $R_x$~(a) and $R_y$~(b) for different horizontal and vertical excitations respectively, averaged across all the BPMs of the KARA model. The error-bars represent one standard deviation from the average values. The theoretical values are shown in black, and the uncertainty due to statistical errors in the estimation the RMS energy spread, $\sigma_{\delta}$, is shown in light green.}
\end{figure}

The dependence of the horizontal chromatic ratio $R_x$ on the number of turns $N$ and on the initial excitation amplitude, can be visualized in Fig.~\ref{sims:FIG9}, while the same dependence for the vertical chromatic ratio $R_y$ are shown in Fig.~\ref{sims:FIG10}. The values represent the average across all the BPMs of the KARA model, and the error-bars represent one standard deviation from the average. Both figures contain the values that are expected from the model, by directly using the simulation parameters in the r.h.s. of Eq.~\eqref{sec:Anal:eq7}, for each initial excitation. An average value is extracted which is shown in black, and one standard deviation around this value is shown in green. This step is taken since the RMS energy spread $\sigma_{\delta}$ depends on the initial longitudinal distribution of the particles, which is generated by using different seeds for each transverse excitation, thus creating variance in the values of the RMS energy spread $\sigma_{\delta}$. It is noted that the \emph{standard error} of the RMS energy spread $\sigma_{\delta}$, between the samples of different initial excitation, is at the range of $1~\%$ for both transverse planes.

The horizontal chromatic ratio $R_x$ exhibits a significant TbT spread among the different excitations of the beam, which cannot be explained by the uncertainty in the estimation of the RMS energy spread $\sigma_{\delta}$. This shows that the main horizontal amplitudes $A_{0x}$ dominate in magnitude over the sum of the first order chromatic sidebands $A_{1x}+A_{-1x}$. In other words, while the main amplitudes increase with a certain step for each case, the chromatic sidebands exhibit a smaller increase for the same cases. 
Moreover, the TbT decay of the horizontal chromatic ratios indicate that the damping rate of the first order chromatic sidebands $A_{1x}$ and $A_{-1x}$ dominates over the damping rate of the horizontal main amplitude $A_{0x}$. While the cases of $2-4~\sigma_x$ excitations produce a very similar chromatic ratio at $N=180$ turns, the $1~\sigma_x$ case exhibits an offset of around $0.02$ from that value. At the same time, the $1~\sigma_x$ case exhibits the least steep decay due to decoherence, which falls outside the expected band of values for this specific number of turns. This suggests that for the current simulations, the $1~\sigma_x$ excitation is not enough for precise chromaticity estimations. The curves of $2~\sigma_x$ and $3~\sigma_x$ initial excitation, exhibit similar trends, albeit a small separation is visible after $N=360$ turns, or after about six synchrotron periods. The $4~\sigma_x$ excitation undergoes a sharp decrease in the very first turns, due to non-linearities, and then quickly decreases in magnitude. This behaviour points to the fact that the available number of turns for precise chromaticity measurements, is limited for this case.

Concerning the vertical plane, the chromatic ratio values $R_y$ remain inside the expected margin of values during TbT analysis, while their distribution, with respect to the initial vertical excitation, is more homogeneous compared to the horizontal plane, i.e. the curves are closer with each other. This characteristic allows for accurate measurements of the vertical chromaticity by using the proposed TbT method. However, it is clear that the $40~\sigma_y$ case demonstrates a chromatic ratio which appears at a distance of around $0.005$ below the average value of the rest of the values. This anomaly can be attributed to the dominance of the $A_{0y}$ term over the sum term $A_{1y}+A_{-1y}$ in Eq.~\eqref{sec:anal:eq13} for this particular case, which drives the chromatic ratio to lower than expected values. As a matter of fact, the analysis shows that even for the $50~\sigma_y$ the chromatic sidebands are larger than expected due to the combination of high chromaticity values and strong initial excitation. The result is that the $50~\sigma_y$ case overlaps with the $30~\sigma_y$ case. In addition, the large vertical chromaticity induces a faster decrease of the vertical chromatic sidebands, compared to the horizontal ones, with a rate that is similar for every excitation. In general, it can be concluded that, due to the difference in chromaticities, the TbT behaviour of the vertical chromatic sidebands $A_{\pm1y}$ dominate over the main amplitudes $A_0y$, while the opposite is true for the horizontal amplitudes $A_{\pm1x}$ and $A_{0}$.

\vspace{2pt}
\subsection{Chromaticity Estimations\label{sims:sec:chrom}}

The findings of the previous section lead to the conclusion that the chromaticity estimations, via the proposed method, depend on the number of turns $N$ of the TbT BPM signal. An immediate result is that one should use the smallest number of turns $N$ for the estimation of the chromatic properties of the beam, in order to avoid the effects of decoherence. At the same time, the number of turns $N$ must allow for the observation of chromatic sidebands with enough precision i.e. the BPM signal should contain several synchrotron periods. In this section, the goal is to investigate the output of the proposed method with respect to different initial excitations.

\subsubsection{Exact solution VS Approximation}

\begin{figure}[!htb]
  \centering
  \includegraphics[width=\mysize\textwidth]{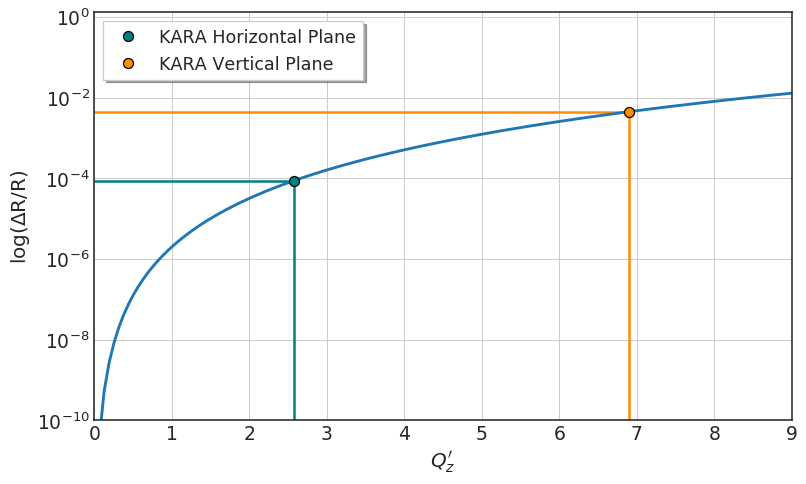}
  \caption{Relative computational error in the approximation of the analytical formula for the chromaticity estimations. The error is presented against the chromaticity $Q_z'$, and it is computed by taking into account the actual values of the synchrotron tune $Q_s$ and the RMS energy spread $\sigma_{\delta}$ of the KARA simulation model. The relative errors at the horizontal chromaticity value of the KARA model is given in green and at the vertical chromaticity in orange.}
  \label{sims:FIG1000}
\end{figure}

The proposed method presents two simple equations for estimating chromaticities from TbT data. By knowing the chromatic ratios, the synchrotron tune, and the RMS energy spread, one can obtain the chromaticities of the accelerator by solving directly Eq.~\eqref{sec:Anal:eq7} or by approximating the solution with Eq.~\eqref{sec:Anal:eq11}. Depending on the aforementioned beam dynamics parameters, the error of the approximation varies, and it is left to the user and the required error margins to decide which formula to use for the estimation of chromaticities. 

For the sake of completeness, the logarithm of the relative computational error $\Delta R / R$, with $R$ being the chromatic ratio predicted by Eq.~\eqref{sec:Anal:eq7}, and $\Delta R$ the error between this prediction and the approximation from Eq.~\eqref{sec:Anal:eq11}, is shown in Fig.~\ref{sims:FIG1000}. The error is shown in blue with respect to an increasing chromaticity $Q_z'$, and the computational error for the horizontal and vertical chromaticities in the KARA simulations are shown in green and orange respectively. 

As it is testified from the trend of the error curve, lower chromaticities can be inferred by using the approximation formula directly, without incorporating a large computational error. Larger chromaticities favor the employment of the exact formula, if the error becomes too important for the intended application of the method. Clearly, due to the large KARA vertical chromaticity, the computational penalty on using the approximation is a bit less than $10^{-2}$, relative to the estimated vertical chromatic ratio $R_y$. The same error is at around $10^{-4}$ for the horizontal case. 

It should be noted that if the suggested method is used experimentally without having a rough estimate for the actual chromaticity in order to assess the magnitude of the error, one could use directly the approximation in order to obtain the chromaticity and then decide if the error levels demand the re-calculation of the chromaticity estimates via the exact formula.

\subsubsection{Accuracy}

 Based on the results of the previous section, chromaticity is estimated by the use of the exact formula, Eq.~\eqref{sec:Anal:eq7}, in order to minimize unwanted sources of systematic errors. However, using the approximation of Eq.~\eqref{sec:Anal:eq11} would not significantly disturb the horizontal chromaticity estimations, as opposed to the vertical plane, where the exact formula should yield more accurate results. 

 The value of the synchrotron tune $Q_s$ used in the analysis, are obtained for each initial excitation, BPM, and number of turns $N$, from the Fourier spectra of the beam, as presented in Section~\ref{subsub:qs}. The number of turns $N$ correspond to a range of three to six synchrotron oscillations $\tau_s$, where $\tau_s=63$ turns.

\begin{figure}[!htb]
\centering
\subfloat[Relative error in the horizontal chromaticity measurements, with respect to the initial horizontal excitation of the beam. The error-bars indicate one standard deviation from the average value of all the BPMs.\label{sims:FIG16}]{\includegraphics[width=\mysize\textwidth]{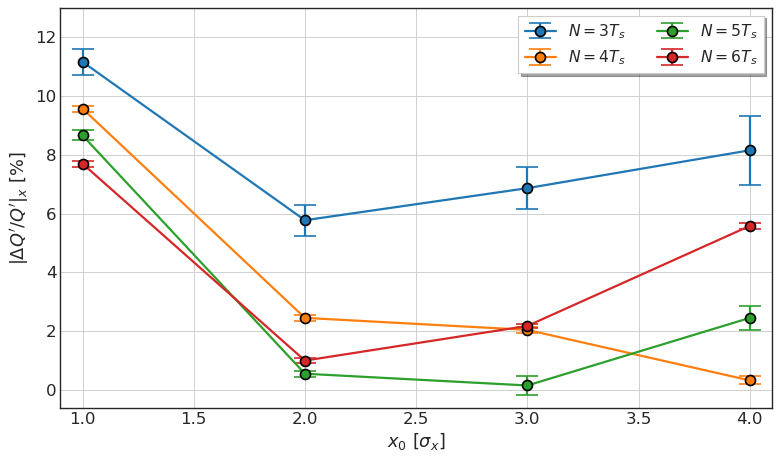}} \\
\subfloat[Relative error in the vertical chromaticity measurements with respect to the initial vertical excitation of the beam. The error-bars indicate one standard deviation from the average value of all the BPMs.\label{sims:FIG17}]{\includegraphics[width=\mysize\textwidth]{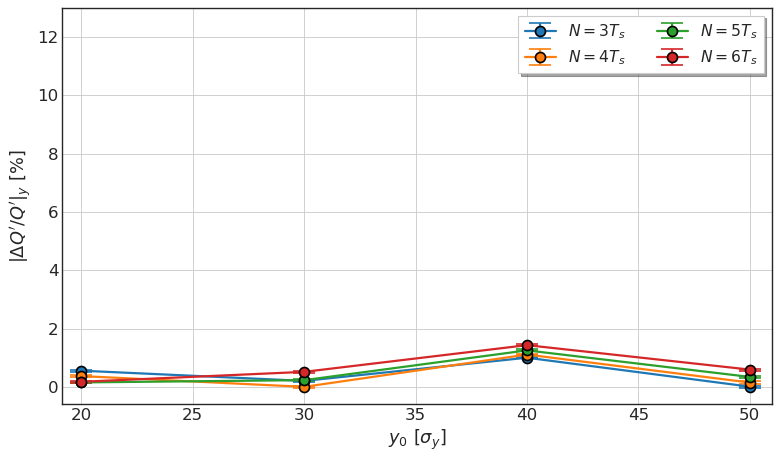}} 
\caption{Absolute relative error of the horizontal~(a) and vertical~(b) chromaticity measurements for the simulations with the KARA accelerator model. Each chromaticity estimate is the average from all BPMs. The relative errors are shown as a percentage of the reference value, against the initial excitation amplitudes of the simulation. The error-bars measure one standard deviation from the average value. The different colors refer to the number of turns $N$ that were used for the measurements, which are multiples of the synchrotron period $\tau_s = 63$ turns.}
\end{figure}

  By using the chromaticities $Q_{x0}'$ and $Q_{y0}'$ that are calculated from \emph{MAD-X} as reference values, the absolute normalized error $|\Delta Q'/Q'|_x=|Q'_x-Q'_{x0}|/Q'_{x0}$ of the horizontal chromaticity $Q_x'$ is presented in Fig.~\ref{sims:FIG16}, with respect to the initial excitation amplitude $x_0$, in units of the RMS horizontal size $\sigma_x$. The absolute error of the vertical chromaticity $|\Delta Q'/Q'|_y = |Q'_y-Q'_{y0}|/Q'_{y0}$, with respect to the vertical excitation in units of RMS vertical size $\sigma_y$, are shown in Fig.~\ref{sims:FIG17}. For both planes, the measurements are given as a percentage of the reference value, and they refer to the average values from all the BPMs, with the error-bars representing one standard deviation from the average value.

As it has been already discussed, the $1~\sigma_x$ case exhibits the least accurate chromaticity estimations, with a maximum normalized error of just below $12~\%$. Integrating the analysis over a longer number of turns seems to improve the error, but not significantly. However, if the initial excitation amplitude $x_0$ increases, the value of the chromaticities converge around the expected value with a normalized error $|\Delta Q'/Q_x'|$ of around $1~\%$ for the $2~\sigma_x$ excitation and a number of turns $N$ of six synchrotron periods (red curve). As a matter of fact, further increase of the excitation amplitude and appropriate choice of the number of turns $N$ leads to even smaller errors, which correspond to below $0.1~\%$ of the reference value of the horizontal chromaticity $Q'_x$. Such a case is presented for the $3~\sigma_x$ and $4~\sigma_x$ excitations, and for a number of turns equal to $N=5~\tau_s$ (green curve) and $N=6~\tau_s$ (red curve) respectively.

Concerning the vertical chromaticity $Q_y'$, the estimations are very close to the value provided by the model, exhibiting overall relative errors of below $1~\%$ for a number of turns equal to $N=5-6~\tau_s$ and below $0.1~\%$ for $N=3-4~\tau_s$. The higher accuracy per excitation, with respect to the horizontal plane, is explained by the larger vertical chromaticity, which makes the spectra more resolvable during frequency analysis, and the distribution of the vertical chromatic ratios which falls inside the expected margin of values, as it is presented in Fig.~\ref{sims:FIG10}. The bump that is observed at $40~\sigma_y$, and sets the error just below $2~\%$ for a number of turn $N=5-6~\tau_s$ and around $1~\%$ for $N=3-4~\tau_s$, is due to the observed deviation of the chromatic ratio for that particular case from the expected value, which is also presented in Fig.~\ref{sims:FIG10}.

For both transverse planes, the spread of the chromaticity estimations across the BPMs of the KARA model is observed to be below $0.05~\%$ of the average value. This observation confirms one of the assumptions of the method that the chromaticity measurement should be independent of the location in the ring.

An immediate conclusion from the results of these simulations is that accurate chromaticity estimations, by using the Fourier sidebands on the main betatron lines, are possible, if the RMS energy spread of the beam and the synchrotron tune are known. However, care has to be taken for the choice of initial excitation of the beam, as it is found to play a role in the estimations of chromaticity. Concerning the current simulations, the dependence is more pronounced for the horizontal plane, where a minimum of the error in the chromaticity estimations can be achieved by a proper choice of the number of turns $N$, and the initial amplitude of the beam.

\subsection{Chromatic beta-beating estimations}

The process of measuring chromatic beta-beating in circular accelerator are similar to the measurement of chromaticity~\cite{Calaga:2010zzc}: The RF system is varied in order to alter the energy of the beam around the reference value, and the response of the beta function is then measured by other experimental methods. A fit of the beta function measurements to the known RF energy deviations estimates the chromatic beta-beating of the lattice. 

The proposed method in this paper, simply utilizes an initial transverse excitation of the beam, in order to produce coherent betatron oscillations. With knowledge of the chromaticity $Q'_z$ and the synchrotron tune $Q'_s$, the chromatic beta-beating can be inferred from Eq.~\eqref{sec:Anal:eq12}. Note that, if chromaticity is not known, it can be measured directly with Eq.~\eqref{sec:Anal:eq7} or Eq.~\eqref{sec:Anal:eq11}.

Concerning the current simulations, chromatic beta-beating is estimated by measuring the chromatic sidebands with \emph{PyNAFF}, and by using the model values of the synchrotron tune $Q_s$ and the chromaticity $Q'_z$, where $z=x, y$. According to the conclusions of Sec.~\ref{sims:sec:sb}, a number of turns $N$ equal to 5~synchrotron periods is used for the analysis. It is found that, for both planes, there is no significant change in the estimated chromatic beta-beating with respect to the initial excitation. The weak dependence on the initial excitation can be explained by the fact that in the chromatic beta-beating estimations, only the chromatic sidebands $A_{\pm1}$ are used, while for the chromaticity measurements, the main amplitude $A_0$ of the beam is needed, which depends strongly on the initial excitation.

\begin{figure}[!htb]
\centering
\subfloat[Horizontal chromatic beta-beating for the 35 BPMs of KARA. The model values are superimposed on the results.\label{sims:dbbx_new}]{\includegraphics[width=\mysize\textwidth]{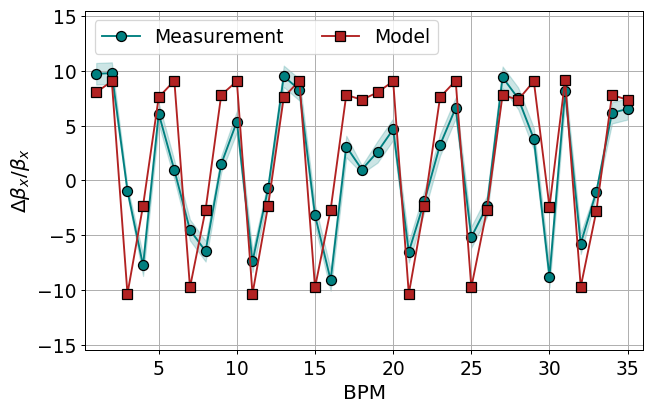}} \\
\subfloat[Vertical chromatic beta-beating for the 35 BPMs of KARA. The model values are superimposed on the results.\label{sims:dbby_new}]{\includegraphics[width=\mysize\textwidth]{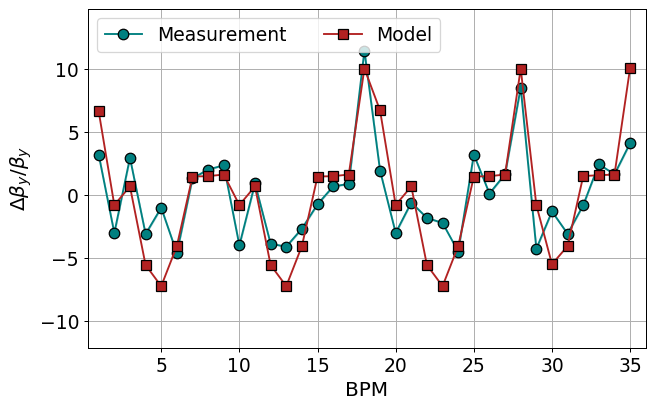}} 
\caption{Chromatic beta-beating measurements using the proposed method, for each BPMs of the KARA model. Horizontal chromatic beta-beating is shown in~(a), averaged across the results from $1-4~\sigma_x$ simulations. Vertical chromatic beta-beating is shown in~(b), averaged across the results from the $20-50~\sigma_y$ simulations. The light green band across the measurements signifies one standard deviation from the average. The model values are shown in red for both planes.}
\end{figure}

The estimated horizontal chromatic beta-beating averaged across $1-4~\sigma_x$ initial excitations is shown in Fig.~\ref{sims:dbbx_new}, while the vertical chromatic beta-beating, averaged across $20-50~\sigma_y$ initial excitations, is shown in Fig.~\ref{sims:dbby_new}. The light green band around the measurements defines one standard deviation from the average. The response of the model beta functions are obtained by using the \emph{PTC} module of \emph{MAD-X}, and it is superimposed on the results. 

It is important to note that the method presented in this paper does not differentiate between negative and positive solutions i.e. the sign of the chromatic beta-beating is ambiguous, which means that only absolute values of chromatic beta-beating can be meaningfully compared with the model expectations. In order to perform a more realistic comparison between the two, the average values are subtracted to center the waves around zero, and the phase of the beta-beating wave of the model is introduced in the measurements, by first determining the absolute chromatic beta-beating with Eq.~\eqref{sec:Anal:eq12}, and then multiplying with the \emph{sign} of the model values. In the absence of reference measurements from the model, this operation would be of course impossible, and only absolute values of the chromatic beta-beating could be used.

The results from the horizontal plane exhibit a good agreement of with the expected chromatic beta-beating. The maximum chromatic beta-beating, estimated by both the model and the measurements, is around~$\frac{\Delta\beta_x}{\beta_x}=10$ for the current simulated KARA optics. Some notable discrepancies between model and measurements do exist, such as at BPM $3$ or in the sections of BPMs $17$ to $20$, and using a different number of turns did not improve the convergence to the model values. On the other hand, the small variance of the measurements in the selected initial excitations, does not explain the offset either.
For the vertical plane, the results from the proposed method are also in good agreement with the expectations from the model. The maximum vertical chromatic beating is found at a value of ~$\frac{\Delta\beta_y}{\beta_y}=10$, while the reproducibility of the results, within the sample of the initial vertical excitations, is excellent. 

Note that, as expected from theory, the results from the proposed method return a well defined beta-beating wave for both planes, with a frequency of almost twice the corresponding betatron tune. The measurements are summarized in Table~\ref{table2}, where the measurements and model values of the RMS chromatic beta-beating is given, for each transverse plane. The RMS values that are quoted in the table are calculated as averages from the samples of RMS chromatic beta-beating of the KARA BPMs for each initial excitation. The quoted error represents one standard deviation from the measurements. The proposed method underestimates the expected RMS chromatic beating by a factor of $18~\%$ for the horizontal plane and a factor of $24~\%$ for the vertical plane, while the uncertainty in the vertical RMS chromatic beta-beating is $10$ times smaller than the horizontal one due to the low impact of non-linearities in the vertical plane. 

\begin{table}[htb!] 
  \small
  \centering
  \caption{RMS Chromatic beta-beating estimations from the KARA model and the proposed method. The RMS values from the proposed method are given as an average of the set of different initial excitations of the simulated particles. The initial excitations are quoted in the brackets. The measurements error is one standard deviation from the average value.}
  \begin{tabular}{l|l|l}
      \toprule
 RMS Chromatic beta-beating &  \textbf{Model} & \textbf{Measurements} \\
\toprule
       Horizontal~$[1~\sigma_x-4~\sigma_x]$  &  7.48 &  $6.11 \pm 0.20$     \\ 
       Vertical~$[20~\sigma_y-50~\sigma_y]$ &  4.70 &  $3.57\pm0.02$     \\
 \toprule
  \end{tabular}
  \label{table2}
\end{table}

\section{Experimental measurements at KARA\label{sec:Meas}}

\subsection{Proof of concept\label{subsec:exp:concept}}

The proposed method for estimating chromaticity through spectral analysis of TbT data is tested experimentally at the KARA light source. A total of $n_b=110$ bunches can be injected in the ring, with a range of nominal beam currents at $I_b=90-120$~mA. The energy of the particles at flat-top is $E=2.5$~GeV and for the nominal KARA optics, the synchrotron tune is around $Q_s=0.012$ while the measured bunch length for the nominal machine settings at KARA is $\sigma_z=1.3\cdot10^{-2}$~m, which can be used to estimate the RMS energy spread at $\sigma_{\delta}=8.9\cdot 10^{-4}$.

\begin{figure}[!htb]
\centering
\subfloat[Horizontal TbT data, after excitation of the injection kicker in a range of currents, which is shown in the legend.\label{exp:FIG18}]{\includegraphics[width=\mysize\textwidth]{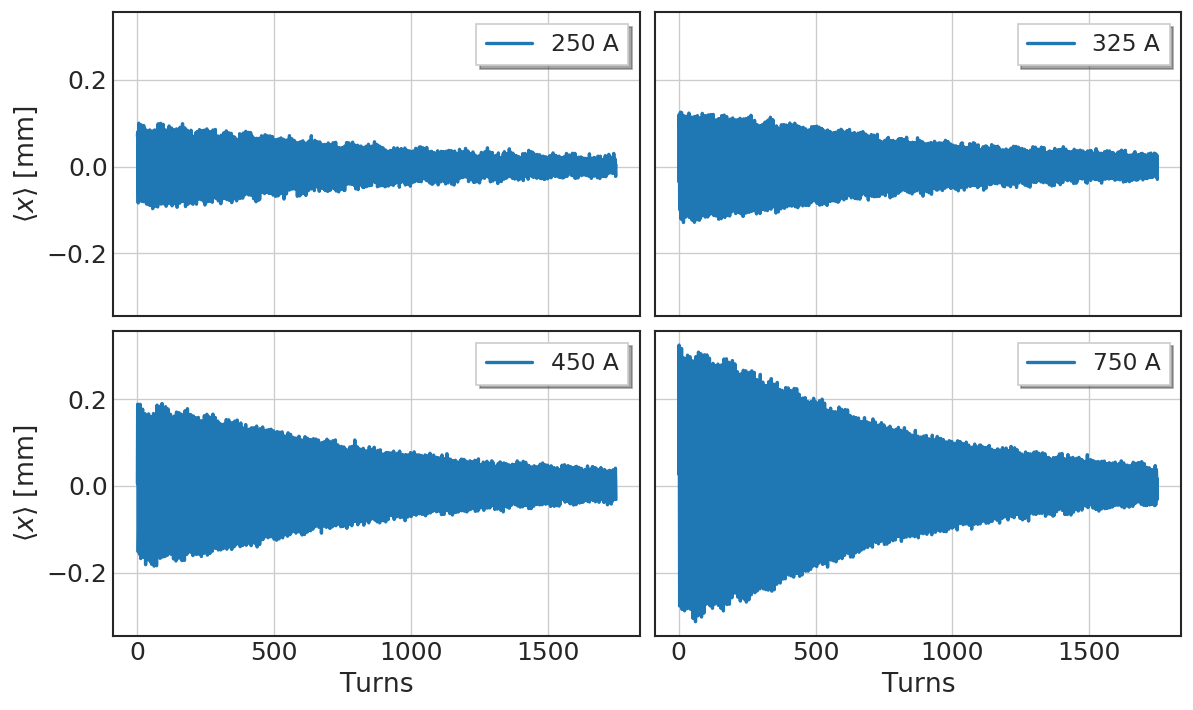}} \\
\subfloat[Vertical TbT data, generated from betatron coupling, following the horizontal excitation of the beam. \label{exp:FIG19}]{\includegraphics[width=\mysize\textwidth]{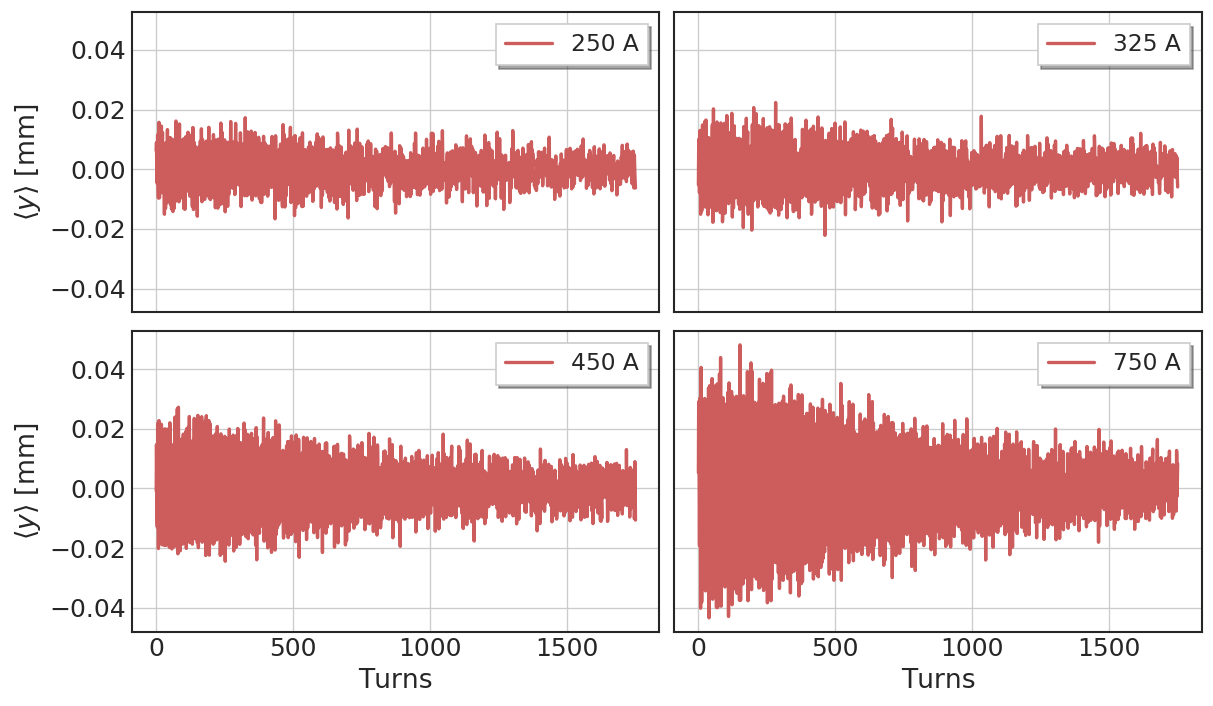}} 

\caption{Experimental TbT data at KARA light source. The horizontal oscillations are presented in (a) and the vertical oscillations in (b). The TbT data are generated by the KARA injection kicker and the current at the kicker for each case is shown in the legend of the graphs.}
\end{figure}

At the time of the measurements only $M=35$ BPMs were available for recording the beam's centroid. The TbT data are generated by the use of the injection kicker, which can induce horizontal betatron oscillations. The vertical excitation is transferred to the beam by virtue of betatron coupling, which naturally results in vertical TbT signals with a much lower Signal-to-Noise Ratio (SNR). The data are gathered for around $N=2000$~turns and the kick from the magnet lasts for about $N=8$ turns or about $2.5$~ms. For the presented experimented, a range of kicker currents from $100$~A to $800$~A is used, in order to generate initial excitations of different amplitudes. A sample of the experimental horizontal TbT data can be visualised in Fig.~\ref{exp:FIG18}, for different magnitudes of the kicker current, which give a maximum RMS oscillation amplitude of $\sqrt{\langle x^2\rangle}=0.2$~m for kicker current of $I_k=750$~A. The vertical TbT data for the same kicker configurations are shown in Fig.~\ref{exp:FIG19}, where the loss of signal purity is evident from the random spikes arising from the low oscillation amplitudes. At a kicker current of $I_k=750$~A, the RMS vertical oscillation amplitude $\sqrt{\langle y^2\rangle}$ is 10 times smaller than the horizontal one. As a result, larger systematic errors are expected in the vertical chromaticity measurements.

\subsubsection{First chromaticity measurements with the new method\label{subsec:exp:meas}}

\begin{figure}[!htb]
  \centering
  \includegraphics[width=\mysize\textwidth]{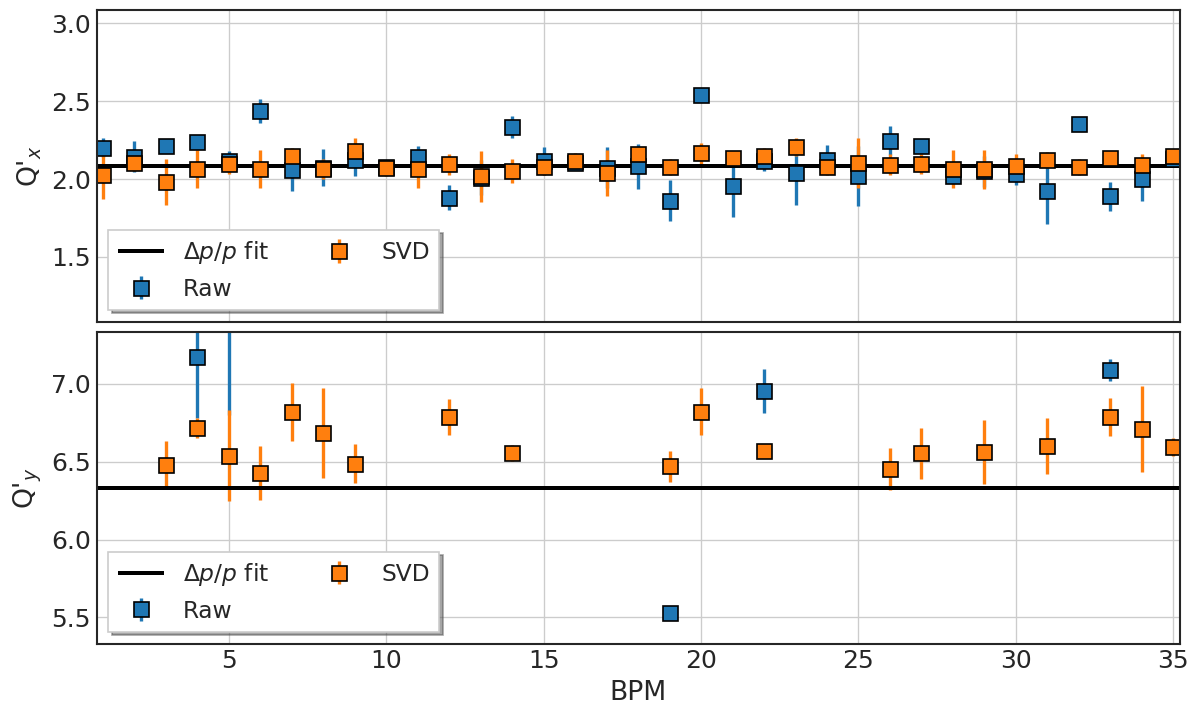}
  \caption{Horizontal (top) and vertical (bottom) chromaticity at KARA, by using the proposed method (blue and orange) and the traditional method (black line). The blue markers correspond to measurements without noise filtering of the TbT data, while orange ones, to data that have been filtered with the SVD method.}
  \label{exp:FIG20}
\end{figure}

The first chromaticity estimations from TbT data, based on the application of the proposed method for a number of turns equal to 5 synchrotron periods, is shown in Fig.~\ref{exp:FIG20}, for the horizontal chromaticity $Q'_x$ (top) and vertical chromaticity $Q'_y$ (bottom), with respect to each KARA BPM. The chromaticity estimations from the raw experimental data are shown with blue markers, while the estimations from the same TbT data, but after post-processing with Singular Value Decomposition (SVD)~\cite{Wang:1999tt} analysis in order to increase SNR, are shown in orange. Both markers correspond to the average value from 10 injections of bunches, and the error-bars represent one standard deviation from the average. In addition, the method is benchmarked against the traditional method for chromaticity measurements that is used at KARA ring, i.e. with the RF sweep. More specifically, for the generation of the coherent betatron oscillations, the current at the horizontal injection kicker is set to the nominal value ($I_k=450$~A), the RF frequency is changed over a range of values, and the betatron tune measurements are performed with the \emph{bunch-by-bunch} (BBB) feed-back system~\cite{Hertle:IPAC2014-TUPRI074}.

For the horizontal plane, it is evident that the proposed method can be used for on-line chromaticity measurements, as the estimations agree very well with the RF sweep method. Even for the raw data the agreement is impressive, while the outliers are absent in the same measurements with the SVD filtered data. The statistical error from 10 consecutive shots is at around $3~\%$ while the agreement with the RF sweep method is below $1~\%$ for the SVD measurements.

Concerning the vertical chromaticity measurements, it is clear that not all BPMs could yield Fourier spectra with detectable chromatic sidebands. However, for the SVD filtered data, the population of the successful BPMs increases due to the improvement of SNR, yielding measurements which slightly overestimate the RF sweep results. In any case, the vertical chromaticity is estimated with an accuracy of around $9~\%$ for the SVD measurements.

Similar measurements can be employed, by computing the BPM average with respect to the excitation kicker current, in order to quantify the effect of the initial excitation in the TbT chromaticity estimation.

\begin{figure}[!htb]
\centering
\subfloat[Horizontal Chroma VS Kick.\label{exp:FIG24}]{\includegraphics[width=\mysize\textwidth]{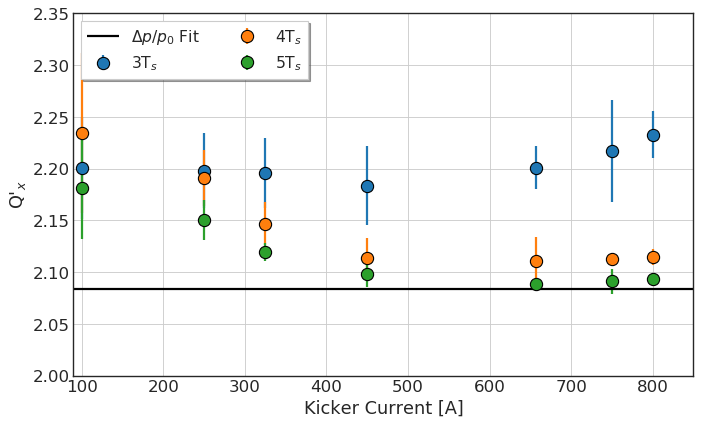}} \\
\subfloat[Vertical Chroma VS Kick. \label{exp:FIG36}]{\includegraphics[width=\mysize\textwidth]{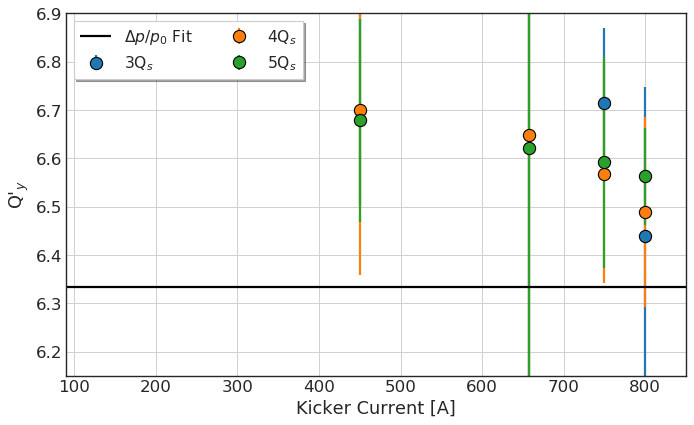}} 

\caption{Estimation of chromaticity at KARA by using the proposed method, with respect to the current of the excitation kicker. The synchrotron period is $\tau_s=\frac{1}{Q_s}=90$~turns. The measurements are produced from analysis of several synchrotron periods, which correspond to $3~\tau_s$ (blue), $4~\tau_s$ (blue) and $5~\tau_s$ (green). The error-bars indicate one standard deviation from the average value all the BPMs at KARA. The measurements from the traditional RF sweep method are superimposed with black lines.}
\end{figure}

Such a measurement is graphically presented in Fig.~\ref{exp:FIG24} for the average horizontal chromaticity from all BPMs, with respect to the kicker current $I_k$, and for a number of turns equal to three (blue), four (orange) and five (green) synchrotron periods. The error-bars correspond to one standard deviation from the average value. The vertical amplitude-dependent chromaticity measurements are shown in Fig.~\ref{exp:FIG36} for the same number of turns and color code as before.

Clearly, as the kicker current increases, the horizontal chromaticity estimations converge to a value which agrees very well with the measurement provided by the RF sweep method, with five synchrotron periods yielding the most accurate and precise results. More specifically, for the aforementioned number of turns, the horizontal chromaticity measurements converge to the expected value at around $I_k=650$~A, while at $I_k=100$~A the error in accuracy is estimated at around $5~\%$. Note that, when the least number of turns is used, the error between the two methods increases after the current of the kicker is set to $I_k=420$~A.

The impact of the SNR of TbT beam position signal is visible at the bottom plot, where the vertical chromaticity is only reproducible for an excitation kick of $I_k\geq425$~A. The choice of a large number of turns seems to improve the measurement error, which reduces significantly at $I_k=800$~A. The reproducibility of the vertical chromaticity measurements is lower than the horizontal plane, due to the inexistence of a pure vertical kick.



The particular dependence of the accuracy of the proposed method, for an increasing excitation amplitude, is also observed in the simulations of Sec.~\ref{sec:Sims}, Fig.~\ref{sims:FIG16} and Fig.~\ref{sims:FIG17}. As an immediate conclusion, the excitation amplitude should be large enough as to induce chromatic sidebands which can be accurately resolved. Empirically, such a configuration would be an excitation amplitude that allows for around $4-5$ synchrotron periods, before it reaches $50~\%$ of the initial amplitude of the centroid at the BPMs. On the other hand, a limit to the maximum excitation should also exist due to non-linearities, but it was not observed until the maximum current of the horizontal kicker.

\subsection{The CLIC Superconducting Wiggler\label{subsec:clic}}

The KARA light source has been recently selected to commission the new prototype of the Compact Linear Collider (CLIC)~\cite{CLIC:1500095} Superconducting (SC) damping wiggler~\cite{Bernhard:IPAC2016-WEPMW002}, which will be used at the Damping Rings~\cite{Papaphilippou:IPAC2012} of CLIC in order to cool-down the electron and positron beams, i.e. reduce the initially large transverse emittances. The basic parameters of the CLIC SC wiggler, which define the linear dynamics, are summarized in Table~\ref{table3}.

\begin{table}[htb!] 
  \small
  \centering
  \caption{Basic parameters of the CLIC Superconducting Damping Wiggler prototype, commissioned at the KARA light-source }
  \begin{tabular}{|l|l|}
 \hline     
 Max. on-Axis Magnetic field $B_w$~[T]& 2.90 \\
 \hline
 Period length $\lambda_w$~[mm]    & $51.40$\\
 \hline
 Total length $L$~[m]    & 1.85\\
 \hline
 Horizontal beta function $\beta_x$~[m]   & 18.96\\
 \hline
 Vertical beta function $\beta_y$~[m]   & 2.17\\
 \hline
  \end{tabular}
  \label{table3}
\end{table}

Since an insertion device, as powerful as the CLIC SC wiggler, can potentially influence the beam dynamics of the KARA ring, a series of measurement campaigns have been deployed for the characterisation of the linear and non-linear beam dynamics' response in the presence of the wiggler~\cite{Bernhard:2013wza,Gethmann:IPAC2017-WEPIK068,Papash:IPAC2018-THPMF070}. 

The first set of measurements during the commissioning of the CLIC SC wiggler at KARA, focuses on the evaluation of the transverse betatron tunes and chromaticities for various magnetic fields of the wiggler. 

Due to the symmetries of the magnetic field components at the wiggler, it acts as a \emph{focusing quadrupole} in both planes~\cite{Venturini:2003ar}. For the present study, the following points from theory are taken into consideration: 

\begin{enumerate}[label=\roman*)]
  \item A vertical focusing component is expected, due to the existence of a non-vanishing longitudinal field component at the wiggler, coupled to the trajectory of the beam ("wiggling") along the insertion device. The focusing results in vertical betatron tune-shift which depends on the period and the peak field of the wiggler.

  \item The horizontal focusing in the wiggler is compensated by feed-downs of the sextupolar components. However, a slight horizontal defocusing might be observed, in the case that the feed-downs are larger than the linear focusing of the wiggler. A beam which enters the wiggler with a non-vanishing horizontal position, experiences a gradient in the distribution of the magnetic field at the successive poles of the wiggler. The net effect is a slight horizontal deflection, which can be observed as a horizontal tune-shift. This defocusing depends on the period and the peak field of the wiggler.

  \item The previous horizontal defocusing highlights the existence of sextupolar components at the wiggler magnetic field. In addition, the vertical focusing of a wiggler can have contributions from higher-order multipoles as well. As a result, the wiggler is expected to alter the chromaticity of the ring in both planes, albeit the effect should be smaller in the vertical plane, in the linear regime, i.e. for low values of the wiggler field. Therefore, chromaticity measurements are also important in the commissioning of a wiggler.
\end{enumerate}

The measurement campaign at KARA consisted of gathering and analysing TbT data, while ramping up the magnetic field of the CLIC SC wiggler, in steps of $0.5$~T until the maximum field of $2.9$~T is encountered. The analysis is performed with the \emph{PyNAFF} software~\cite{PyNAFF}, for each of the $M=39$ BPMs available at KARA during these measurements. The stored current during the measurement procedure is around $I_b=100$~mA. The horizontal excitation is delivered through the horizontal injection kicker, while the vertical excitation is induced by virtue of betatron coupling. The linear chromaticities are set to low enough values as to assert the observation of any sextupolar components due to the ramp-up of the wiggler. In order to assess the contribution of the CLIC SC wiggler on the linear beam dynamics, experimental measurements of the CLIC SC wiggler tune-shift and beta-beating are performed, and the results can be inspected in sections \ref{sec:app:wiggler} of the Appendix.

\subsubsection{\label{sec:exp:secsec:rms}RMS Energy spread-shift due to the CLIC Wiggler}

 The dependence of the RMS energy spread $\sigma_{\delta}$ in the magnetic field of a wiggler can be expressed as~\cite{Wiedemann:1979kt}

\begin{equation}
\label{exp:eq4}
  \sigma_{\delta}^2=\sigma_{\delta_0}^2~\frac{1+\frac{L_w}{2\pi\rho_0}\bigl(\frac{\rho_0}{\rho_w}\bigr)^3}{1+\frac{L_w}{2\pi\rho_0}\bigl(\frac{\rho_0}{\rho_w}\bigr)^2}\,\,,
\end{equation}
where $\sigma_{\delta_0}$ is the value of the natural RMS energy spread without the wiggler, $L_w$ is the length of the wiggler, $\rho_0$ is the bending radius of the main dipoles, and $\rho_w$ is the bending radius of the wiggler magnet. The previous relationship depends on the wiggler magnetic field $B_w$ through the conservation of the beam rigidity

\begin{equation}
  \label{exp:eq5}
\rho_w=\frac{B_0\rho_0}{B_w}\,\,,
\end{equation}
where $B_0$ is the magnetic field of the main dipoles. 

The RMS energy spread $\sigma_{\delta}$ of the beam is experimentally measured for each step of the CLIC SC wiggler. The synchrotron light diagnostics at KARA allow for the measurement of the RMS bunch length $\sigma_z$ by using a Hamamatsu streak camera ~\cite{Kehrer:IPAC2015-MOPHA037}. The RMS energy spread $\sigma_{\delta}$ is inferred by 

\begin{equation}
\label{exp:eq3}
\sigma_{\delta}=\frac{Q_s}{a_p R}\sigma_z\,\,,
\end{equation}
where $Q_s$ is the synchrotron tune, $a_p$ is the momentum compaction factor, and $R$ is the radius of the KARA ring. Note that the momentum compaction factor $a_p$ depends also on the magnetic field of the wiggler $B_w$, however the contribution for the CLIC SC wiggler is negligible. The resolution of the streak camera is $\Delta \sigma_z=1.5$ ps, which defines the error of a single RMS energy spread measurement to be $\Delta \sigma_{\delta}=1.4\cdot10^{-5}$. For the current value of the RMS energy spread $\sigma_{\delta}$, the normalised uncertainty of the measurement is around $\Delta \sigma_{\delta}/\sigma_{\delta}=1.6~\%$.

\begin{figure}[!htb]
  \centering
  \includegraphics[width=\mysize\textwidth]{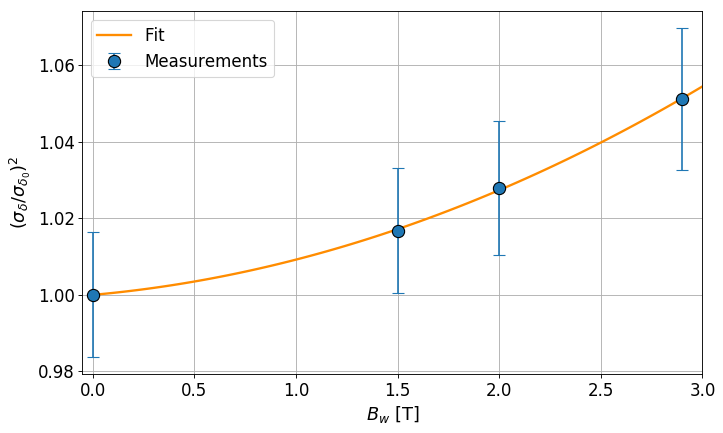}
  \caption{RMS Energy spread measurements (blue markers) from the KARA streak camera, with respect to the magnetic field of CLIC SC wiggler. The error-bars represent one standard deviation from the average. The fit to the theoretical model is shown in orange.}
  \label{sims:FIG23}
\end{figure}

For the current experimental measurements at KARA, the ratio $\bigl(\sigma_{\delta}/\sigma_{\delta_0}\bigr)^2$ is estimated for wiggler fields of $B_w=0.0$~T, $1.5$~T, $2.0$~T and $2.9$~T, due to the unavailability of the streak camera for the intermediate steps of $B_w=0.5$~T, $1.0$~T and $2.5$~T. In order to estimate the RMS energy spread in the missing steps, a non-linear fit of the available RMS energy spread measurements to Eq.~\eqref{exp:eq4} is performed. The results are graphically presented in Fig.~\ref{sims:FIG23}, where the ratio $\bigl(\sigma_{\delta}/\sigma_{\delta_0}\bigr)^2$ is plotted with respect to the CLIC SC wiggler field $B_w$. The experimental measurements are marked with blue, with the error-bars corresponding to the uncertainty of the streak camera measurements, while the aforementioned fit is shown in orange.

From the trend of the measurements, the total increase of the initial RMS energy spread $\sigma_{\delta_0}$ at $B_w=0.0$~T, is around $20\%$ for the CLIC SC wiggler operating at $B_w=2.9$~T.

\subsubsection{Chromaticity-shift due to the CLIC Wiggler\label{sec:exp:chroma_wig}}

In the case of a wiggler with peak magnetic field of $B_w$, additional quadrupolar and sextupolar gradients are superimposed to the magnetic fields of the ring. Furthermore, the wiggler creates additional dispersion, whose average value $\langle D_w\rangle$ scales linearly with the magnetic field of the wiggler $B_w$ as~\cite{Walker:1994mg}

\begin{equation}
    \langle D_w\rangle=-\frac{1}{B_0\rho_0}\frac{1}{k^2}B_w\,\,,
  \end{equation}  
where $k=\frac{2\pi}{\lambda_w}$, $\lambda_w$ is the period of the wiggler, and $B_0\rho_0$ is the magnetic rigidity of the ring. For the KARA ring $\langle D_w\rangle/B_w \propto10^{-6}$~m/T, which gives negligible dispersion at the maximum field of $B_w=2.9$~T, compared to the average dispersion from the main dipoles at the wiggler region $\langle D \rangle \approx 0.4$~m, as given by the KARA model.

In order to establish the presence of sextupolar components in the location of the CLIC SC wiggler, dedicated chromaticity measurements are performed by recording TbT data for around $N=2000$ turns, while ramping up the CLIC SC wiggler from $0$~T to $2.9$~T in steps of $0.5$~T. The analysis is performed with \emph{PyNAFF} by using two methods:

\begin{enumerate}[label=\roman*)]
  \item \textbf{RF-sweep}: During each step, the RF frequency is modulated in order to induce a change in the relative momentum offset of the beam $\delta=\Delta p/p_0$ in the range of $|\delta|\leq 0.5\%$. Chromaticity is estimated from the chromatic response of the betatron tunes for each value of the wiggler field $B_w$. 

  \item \textbf{Chromatic sidebands}: The suggested method for chromaticity measurements is benchmarked against the RF-sweep method. For this method, chromaticity is inferred by inspecting the Fourier spectra of the beam and applying Eq.~\eqref{sec:Anal:eq11}, where the RMS energy spread $\sigma_{\delta}$ estimations are known from previous analysis, see Sec.~\ref{sec:exp:secsec:rms}. The measurements of the chromatic sidebands are performed for around 4 to 5 synchrotron periods. It should be mentioned that the analysis of the chromatic sidebands is performed on TbT data that are gathered while the beam is on the nominal chromatic orbit i.e. for $\delta=0$. For each step of CLIC SC wiggler, three sets of data, where the beam follows the nominal chromatic orbit, are available. This results in fewer statistics than the RF-sweep method, which uses all the available data.
\end{enumerate}

During the initial set-up of the experiment, transverse chromaticity is trimmed to low-enough values ($Q'_x\approx1,~Q'_y\approx4$) by reducing the strength of the lattice sextupoles, in order to observe the effect of the CLIC wiggler. These values are the lower limit that ensure beam stability, since the BBB feedback is not employed during the measurements to avoid distortion of the Fourier spectra. 

\begin{figure}[!htb]
\centering
\subfloat[Horizontal chromaticity, with respect to the magnetic field of the CLIC SC wiggler.\label{exp:FIG25}]{\includegraphics[width=\mysize\textwidth]{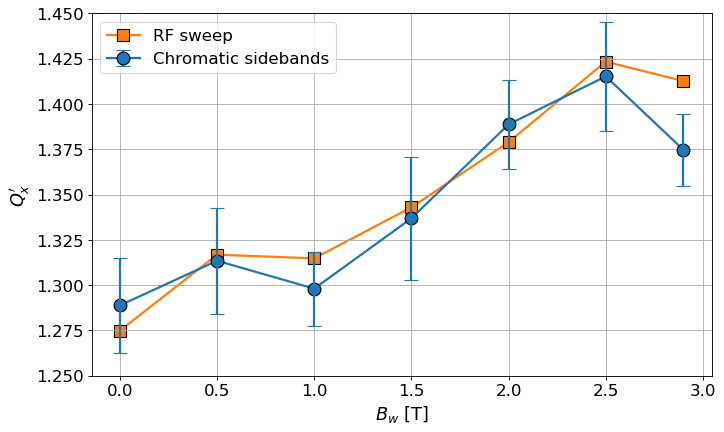}} \\
\subfloat[Vertical chromaticity, with respect to the magnetic field of the CLIC SC wiggler.\label{exp:FIG26}]{\includegraphics[width=\mysize\textwidth]{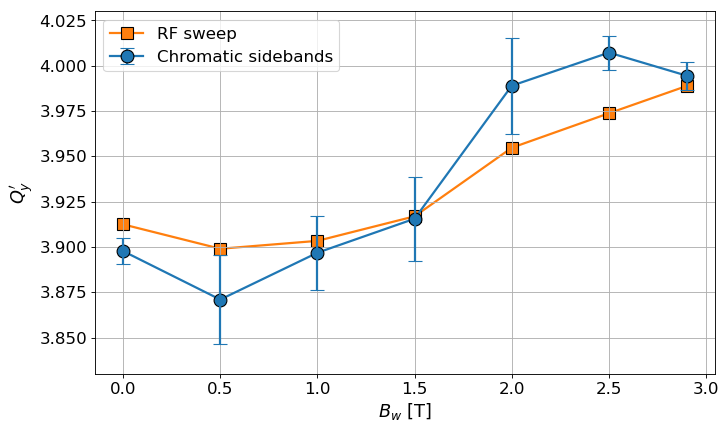}} 
\caption{Experimental measurements of the horizontal (a) and vertical (b) chromaticities by using the proposed method (blue markers) and the standard method of RF sweeping (orange markers). The values from the RF-sweep method correspond to the average from all the 39 BPMs at KARA. The respective measurements of the proposed method quote the average value from all the 39 BPMs and the three available data sets. The error-bars indicate the standard error of the average, where the uncertainty in the RMS energy spread is added in quadrature.}
\end{figure}

The results for the chromaticity measurements with respect to the field of the CLIC SC wiggler $B_w$ are shown in Fig.~\ref{exp:FIG25} for the horizontal chromaticity $Q'_x$ and in Fig.~\ref{exp:FIG26} for the vertical chromaticity $Q'_y$. The estimations with the traditional RF-sweep method are shown in orange and they correspond to the average chromaticity, from all the $M=39$ KARA BPMs. The standard error of the mean is below $10^{-4}$ for both planes, due to the very good Signal-to-Noise Ratio (SNR) and the excellent reproducibility of the betatron tunes for each bunch injection. 
The output from the analysis of the chromatic sidebands with the suggested method are shown in blue, where each marker corresponds to the average value from all the BPMs, and for the three available data-sets, while the error-bars correspond to the standard error of the mean, whereas the uncertainty of the RMS energy spread $\sigma_{\delta}$ measurements has been added in quadrature.

 Concerning the horizontal chromaticity, both methods agree very well, within the margin of error, and both signify the presence of sextupolar components in the CLIC SC wiggler, revealing a positive correlation with the magnetic field of the wiggler. The error-bars for the suggested method is of the order of $10^{-2}$. The trend of the measurements suggests two different regions of horizontal chromaticity increase: a small increase of about $5~\%$ from $0$~T to $1$~T, and a more steep, almost linear increase of about $12~\%$ from $1$~T to $2.5$~T, which slightly drops at the final step of $B_w=2.9$~T. The drop is reported from both methods, and it is possible that ramping the wiggler at the maximum field of $B_w=2.9$~T, can generate additional non-linearities which can slightly perturb the optics. A similar behaviour, albeit of less significance, is observed in the betatron tune measurements presented in Sec.~\ref{sec:app:opt}, where the horizontal tune measurement at $B_w=2.9$~T exhibits a slight focusing at the final wiggler step, with respect to lower fields. Note that magnetic quenches for the CLIC SC wiggler prototype have been reported in~\cite{Bernhard:IPAC2016-WEPMW002} for higher values of the magnetic field $B_w$, limiting the stable operating region to $0\leq B_w\leq2.9$~T.

 As for the vertical plane, there is also a good agreement between the two methods, by considering also the technique for generating the vertical TbT through coupling, which results in a weak vertical TbT signal at the BPMs. Note that in general the uncertainty of the proposed method is smaller than the uncertainty in the horizontal plane, due to the larger vertical chromaticity. However, for some steps of the wiggler field $B_w$, the beam measurements were observed to be less reproducible in the vertical plane, probably due to the simultaneous ramp up and down of the RF system, during the continuous ramping of the wiggler. 

 In any case, the vertical chromaticity measurements reveal a pattern similar to the one observed in the horizontal chromaticity measurements. A pronounced difference is that from $0$~T to $0.5$~T both methods report a drop in the chromaticity, which could be expected from the simultaneous increase of the horizontal chromaticity. However, from $1$~T to $2.5$~T the vertical chromaticity increases, leading to the conclusion that the distribution of the sextupolar components of the CLIC SC wiggler changes with respect the magnetic field.

\subsubsection{Second Order Effects}

During the experimental measurements at the KARA light source, the influence of the CLIC SC wiggler on second order chromaticity $Q''_z$ and chromatic beta-beating $\frac{\Delta\beta_z}{\beta_z}$ is determined, by employing the proposed method, shown in Eq.~\eqref{sec:Anal:eq12}, for the latter, and by using the RF-sweep method for the former. Similar to the linear chromaticity measurements described earlier, the field of the wiggler is ramped up from $0$~T to $2.9$~T, in steps of $0.5$~T. Second order chromaticity is measured by the response of the centroid of the beam to ramping up and down the frequency of the RF cavities, while the chromatic beta-beating is measured after analyzing for the first 5 synchrotron periods the TbT BPM data, for each value of the magnetic field of the CLIC SC wiggler, in order to estimate the amplitude of the chromatic sidebands, $A_{\pm1}$. The values of chromaticity $Q'_z$, measured with the suggested method, are used in Eq.~\eqref{sec:Anal:eq12}.

\begin{figure}[!htb]
\centering
\subfloat[Left axis (green): Horizontal chromatic beta-beating, with respect to the magnetic field of the CLIC SC wiggler. Right axis (red): Horizontal second order chromaticity measurement, for the same values of the magnetic field of the CLIC SC wiggler. \label{exp:dbb_CLIC_hor}]{\includegraphics[width=\mysize\textwidth]{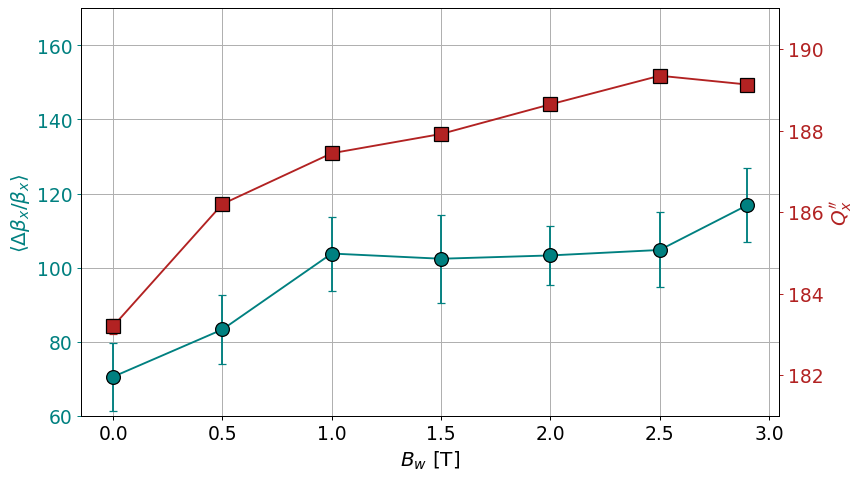}} \\
\subfloat[Left axis (green): Vertical chromatic beta-beating, with respect to the magnetic field of the CLIC SC wiggler. Right axis (red): Vertical second order chromaticity measurement, for the same values of the magnetic field of the CLIC SC wiggler.\label{exp:dbb_CLIC_ver}]{\includegraphics[width=\mysize\textwidth]{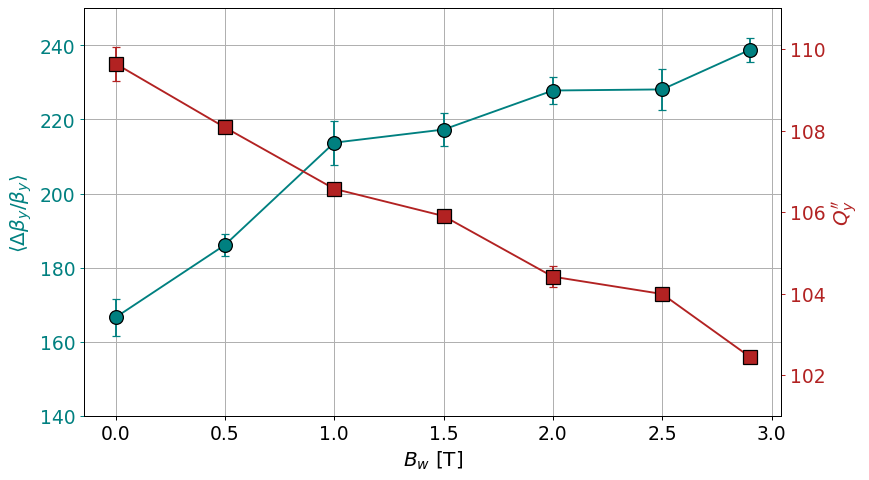}} 
\caption{Correlation of the average chromatic beta-beating (left axis in green), during the ramping of the CLIC SC wiggler, and the corresponding change in second order chromaticity (right axis in red) for the horizontal (a) and vertical (b) planes. The error-bars indicate the standard error from the average, sampled across the KARA BPMs and for three data-sets.}
\end{figure}

The measurements for the horizontal plane are presented in Fig.~\ref{exp:dbb_CLIC_hor}, while the measurements for the vertical plane are shown in Fig~\ref{exp:dbb_CLIC_ver}. Both figures illustrate the evolution of the chromatic beta-beating $\frac{\Delta\beta_z}{\beta_z}$, measured with the suggested method, on the left axis (green color), while the dependence of the second order chromaticity $Q''_z$ on the field of the wiggler is shown on the right axis (red). For both measurements and for both planes, the points are the average measurements, across 3 data sets and 39 BPMs of KARA, and the error-bars represent the standard error of the mean. 

The normalized error for the horizontal chromatic beta-beating is at the range of $\sigma_{\Delta\beta_y}/\Delta\beta_y\approx10~\%$, while for the vertical plane, the same parameter is $\sigma_{\Delta\beta_y}/\Delta\beta_y\approx0.1~\%$. The small error in the vertical plane is explained by the fact that the vertical chromaticity is on average 3 times larger than the horizontal one, which results in more resolvable chromatic sidebands. The errors of the second chromaticity measurements are negligible in both planes, due to the very good reproducibility of the TbT data.

 The trend of the curves for the horizontal plane, exhibits an increase of the chromatic beta-beating of around $100~\%$ with respect to the nominal value at $B_w=0$~T, followed by an increase of the second order chromaticity for around $4~\%$. Note that from $B_w = 1$~T to $B_w=2.5$~T, chromatic beta-beating appears to saturate, while the last point at $B_w=3$~T can be explained by non-linearities, as mentioned in Sec.~\ref{sec:exp:chroma_wig}. 

 The results of the vertical plane indicate that the vertical chromatic beta-beating is resolved with higher confidence by using the proposed method, and that it is found to be also increasing with respect to the CLIC SC magnetic field $B_w$, with a total change of around $50~\%$. However, the growth in the vertical chromatic beta-beating is followed by a decrease in the vertical second order chromaticity $Q''_y$ of about $8~\%$, two times larger than the change in the horizontal second order chromaticity $Q''_x$. 

 Note that the behaviour of second order chromaticity and chromatic beta-beating is fully expected from theory, as it is justified from Eq.~\eqref{exp:eq:chroma_dbb}. An immediate conclusion is that, during experimental measurement, where the reproducibility of the beam is at an adequate level, the chromatic beta-beating can be estimated with the proposed method. Unfortunately, there are no reference measurements in order to compare the results, neither from experiment nor from the model. 


\subsubsection{\label{sec:exp:estimationssex}Estimations of the sextupolar component induced from the CLIC Wiggler}

In the measurements that are presented in the previous section, the first and second order chromaticities, as well as the average chromatic beta-beating of the KARA ring, are found to increase with respect to the field of the wiggler. Even if the observed total growth of these parameters is not severe, the contributions of the CLIC SC wiggler should be estimated in order to have a more systematic analysis.

\begin{figure}[!htb]
\label{fig:ds}
  \centering
  \includegraphics[width=\mysize\textwidth]{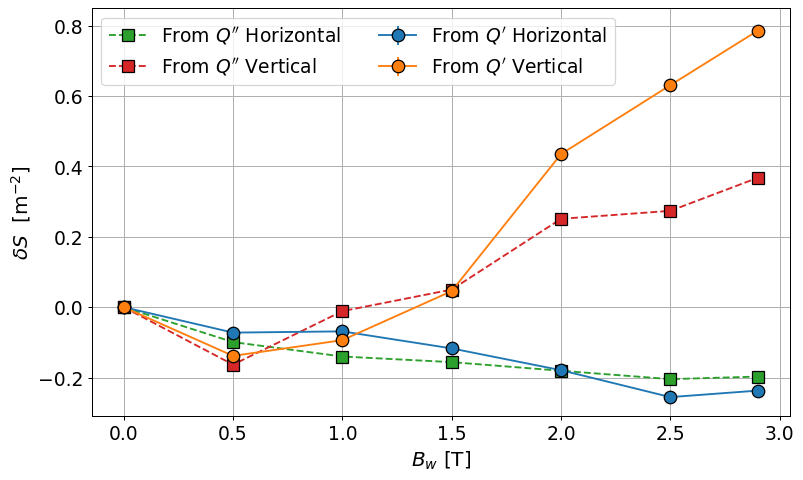}
  \caption{Estimation of the sextupolar contributions during operation of the CLIC SC wiggler, by using two experimental observables: the linear chromaticity $Q'$ (thick lines) and the second order chromaticity $Q''$ (dashed lines). The estimations for the horizontal plane are shown in blue and green, while contributions to the vertical sextupolar component are shown in orange and red.}
  \label{exp:FIG1000}
\end{figure}

By using the experimental measurements of the linear chromaticity $Q'$, the second order chromaticity $Q''$, the average chromatic beta-beating at the region of the wiggler $\bigl\langle\Delta\beta/\beta\bigr\rangle_w$, and the model values of the dispersion $D_w$ and the $\beta_w$ function at the region of the wiggler in Eq.~\eqref{app:eq_ds1} and Eq.~\eqref{app:eq_ds2}, the sextupolar perturbations can be estimated. The results are presented in Fig.~\ref{fig:ds} with respect to the field of the wiggler $B_w$ and for both transverse planes. Note that the values of linear chromaticity $Q'$ and second order chromaticity $Q''$ are taken from the RF-sweep method to minimize the systematic errors in the estimations. The chromatic beta-beating measurement is taken from the suggested method, i.e. from the analysis of the chromatic sidebands of the beam.

The estimations of the horizontal integrated sextupolar components $\delta S_x$ follow a decreasing trend for an increasing wiggler field $B_w$, reaching a value of around $\delta S_x=-0.2$~m$^{-2}$ at $B_w = 2.9$~T. The measurements from both observables, i.e. the linear and second order horizontal chromaticities, agree very well, yielding similar results for the whole range of operation of the wiggler. 
Concerning the measurements for the vertical plane, the sextupolar contribution from the CLIC SC wiggler is initially negative and further decreasing, with values similar  to the horizontal plane. This trend explains the small decrease of the linear vertical chromaticity at $B_w = 0.5$~T, i.e. from the moment that the wiggler was turned on. However, as the field increases, the additional sextupolar component changes polarity and increases as well, which corresponds to the observed increase of the linear vertical chromaticity $Q'_y$.

 The estimated sextupolar contributions by using both vertical first and second order chromaticities agree well for a wiggler field of less than $B_w=1.5$~T, however a discrepancy between the two is obvious for larger wiggler fields. More specifically, at $B_w=0.5$~T the contribution to the vertical sextupolar component from both observables is around $\delta S_y=-0.2$~m$^{-2}$, while at $B_w=2.9T$ the contribution becomes $\delta S_y=+0.8$~m$^{-2}$ if the $Q'_y$ is used, while the same measurement is at about $\delta S_y=+0.4$~m$^{-2}$, if the second order chromaticity $Q''_y$ is employed.  

After inspecting the findings of the previous sections and the assumptions that are taken in the derivations of Eq.~\eqref{app:eq_ds1} and Eq.~\eqref{app:eq_ds2}, the field dependent discrepancy between the results from the two observables suggests that the vertical chromatic beta-beating becomes important for fields $Bw\geq1.5$~T. In addition, the quadrupolar component for the vertical plane is also influenced by the wiggler field, as it is shown in Sec.~\ref{sec:app:tshift} and the observed vertical tune-shift. Note that the two previous points do not hold for the horizontal plane and this is why the results of the two methods agree well with each other. 

By estimating the additional sextupolar components of the wiggle from the linear chromaticity measurements $Q'$ and Eq.~\eqref{app:eq_ds1}, and by knowing the integrated sextupolar components of the KARA ring when the wiggler is off, i.e. around $S_{x,y}=4$ m$^{-2}$~\cite{papash2017high}, the associated changes of the sextupolar component of the KARA ring when the wiggler is operated at $B_w=2.9$~T are

\begin{align*}
\delta S_x/ S_x&\approx5\% \\ 
\delta S_y/ S_y&\approx20\% \,\,,
\end{align*}
for $x,y$ the horizontal and vertical planes respectively. 

With the optics present at KARA during the time of the measurements, the above shifts of the sextupolar components are responsible for the observed relative increase of the chromaticities from $0~$T to $2.9$~T, i.e. $12\%$ for the horizontal plane and $3\%$ for the vertical plane. Even if the relative increase of the vertical sextupolar component is four times larger than the horizontal, the resulting vertical chromaticity shift is four times smaller than the horizontal one. This is explained by the difference between the horizontal and vertical beta functions at the position of the wiggler (Table~\ref{table3}), which once again proves to be indispensable for the minimization of the impact of the CLIC SC wiggler on the KARA optics.

\section{Conclusions\label{sec:Conc}}

Two simple equations are proved and proposed, which can be used to estimate linear chromaticity and chromatic beta-beating, directly from the Fourier spectra of the TbT beam position of an electron beam. The procedure is similar to betatron tune measurements i.e. after applying a transverse excitation to the beam. Such a possibility would add further flexibility to the effort for continuous on-line measurements and control of the beam optics. By using the same method, the RMS energy spread can be also estimated in a TbT manner. 

One of the most important assumptions of the methods is that the distribution of the electron beam is Gaussian, which is almost always the case for electron beams in high-energy circular accelerators, and that the initial excitation of the beam is not too strong as to generate non-linearities which can possibly affect the efficiency of the proposed methods.

Tracking simulations are deployed with the KARA model, where the chromaticity is estimated with the proposed technique. The efficiency of the method is demonstrated, as it shown that chromaticity can indeed be recovered from TbT data. More specifically, the results indicate that two parameters can be used to fine-tune the estimations: the initial excitation amplitude and the number of turns that is used for the frequency analysis. Since decoherence is found to play an important role in the final result, the number of turns should be as small as to allow the generation of synchrotron sidebands in the transverse Fourier spectra. For a powerful frequency analysis tool like \emph{NAFF}, this is usually achieved in four to six synchrotron periods. Similar behaviour is found for the chromatic beta-beating measurements, as it is found that it can be fully recovered in the simulations, via the proposed method. 

The method is also deployed in experimental measurements at the KARA light source, where horizontal excitations are applied to the beam, and the vertical TbT data are generated from betatron coupling. The concept of using the chromatic sidebands for chromaticity estimations is demonstrated with success, and the importance of the signal-to-noise ratio in the beam position signal is highlighted. This is also reflected from the fact that the vertical chromaticity measurements, which are not produced by direct excitation, are less precise and accurate than the horizontal ones. Nevertheless, an important outcome is that the reduction of noise with numerical methods such as the SVD, leads to more precise results. 

In another series of experiments, the method is utilized for the measurement of chromaticity under the influence of the CLIC Superconducting Damping Wiggler. In addition, the impact of the CLIC wiggler on the KARA beam dynamics is characterised qualitatively and quantitatively. The most important outcome of these measurements, is the demonstration of the existence of non-linear fields at the position of the wiggler. The influence of these fields on chromaticity is presented, by using the traditional RF-sweep method and measurements with the proposed methodology. In both transverse planes, the suggested method is benchmarked with success against the traditional technique, and report a simultaneous increase of both transverse chromaticities. The magnitude of the increase however is not severe, and as a result, it is concluded that the operation of the CLIC wiggler at maximum, does not have an important effect on the KARA beam dynamics. 

Moreover, the proposed method is employed in order to estimate chromatic beta-beating and demonstrate its relationship to the wiggler field. At the same time, measurements of the second order chromaticity are performed by using the RF-sweep method. The results indicate an increase of the horizontal chromatic beta-beating, which is followed by an increase of the second order chromaticity. Vertical chromatic beta-beating is found also to be increasing, with a simultaneous decrease of the vertical second order chromaticity.

The additional sextupolar fields due to the operation of the wiggler are estimated by using the previous experimental measurements. The most notable perturbation is reported for the vertical plane with an additional sextupolar component which is four times larger than the horizontal one, but overall $20~\%$ of the integrated sextupolar component of the KARA ring. However, the evolution of the chromatic parameters with respect to the field of the CLIC SC wiggler suggest that the additional fields are not alarming for the stable operation of the KARA ring. 

The proposed method is proven to be reliable and efficient due to the mode of application. However, the dependence of accuracy on the initial excitation and the number of turns used for the spectral analysis, should be always taken into account. Finally, as a future work, a similar methodology could be also developed for proton machines, where the on-line monitoring and control of chromaticity and/or RMS energy spread is of high importance as well.

\section{Acknowledgements}
We would like to acknowledge J. Gethmann, E. Hertle, A.S. Müller and the operators at KARA for assisting with the experiments. This work has been supported by the CLIC project and the Low Emittance Rings (LER) network of EUCARD2 and RULE of ARIES.


\appendix
\section{Insights from theory}

\subsection{Chromaticity\label{sec:app:chroma}}

The \emph{linear} $Q'$ and \emph{non-linear} $Q''$ chromaticities are defined as the betatron tune-shift of a single particle, due to a change of the particle's energy. The energy dependent betatron tune-shift is defined from the well known formula
\begin{equation}
Q(\delta)=Q_0+Q'\,\delta+\frac{1}{2}Q''\,\delta^2+\mathcal{O}(\delta^3)\,\,,
\label{sec:app:Eq1}
\end{equation}
where $\delta=\frac{\Delta p}{p_0}$ with $\Delta p$ the deviation from the reference momentum $p_0$, $Q(\delta)$ is the energy dependent betatron tune, and $Q_0$ is the unperturbed betatron tune, defined by the lattice. 

The linear chromaticity $Q'_{x,y}$ of a circular accelerator can be expressed with respect to the transverse optics as 

\begin{equation}
\label{eq_chroma}
  Q'_{x,y}=\frac{1}{4\pi}\oint_C \beta_{x,y} (s) \bigl[K_{x,y}(s)\mp D_x(s)S_{x,y}(s)\bigr]~ds\,\,,
\end{equation}
where the integration is performed around the circumference of the ring $C$, $\beta_{x, y}(s)$ is the unperturbed (i.e. on-momentum) transverse beta function, $K_{x,y}(s)$ is the quadrupolar focusing gradient, $S(s)$ is the sextupolar focusing gradient, and $D_x(s)$ is the dispersion generated by horizontal bending in the main dipoles.

\subsection{Chromatic beta-beating}
The chromatic beta-beating is defined as the perturbation of the beta function $\beta$, with respect to the momentum offset $\delta$. Expanding the beta function up to first order gives
\begin{equation}
\label{sec:app:eq0}
  \beta(\delta)=\beta_0 +\frac{\partial \beta }{\partial\delta}\,\delta\,\,,
\end{equation}
where $\beta_0$ is the unperturbed beta function. 
The chromatic beta-beating $\frac{\Delta \beta}{\beta}$ is defined in the bibliography~\cite{Fartoukh:1999jk,takao2001higher} as

\begin{equation}
\label{sec:app:eq_1}
  \frac{\Delta \beta}{\beta}=\frac{1}{\beta_0}\frac{\partial \beta }{\partial\delta}\,\,,
\end{equation}
and it can be calculated for both transverse planes by integrating along the whole circular accelerator with the relationship

\begin{align}
  \label{sec:app:eq_dbb}
  \frac{\Delta \beta_{x,y}}{\beta_{x,y}}(s)=&\frac{1}{2\sin(2\pi Q_{x,y})}\oint \beta_{x,y}(s')[K_{x,y}(s') \\
  & \mp D_x(s')S_{x,y}(s')]\cos(2|\phi_{x,y}(s)-\phi_{x,y}(s')| \nonumber \\ 
  &-2\pi Q_{x,y})~ds' \nonumber\,\,,
\end{align}
where $s$ is the path length, $Q_{x,y}$ is the betatron tune, $K_{x,y}(s)$ and $S_{x,y}(s)$ are the quadrupolar and sextupolar integrated magnetic strengths, $D_x(s)$ is the horizontal dispersion and $\phi_{x,y}(s)$ is the phase advance of the betatron oscillations. 

 The expression in~\eqref{sec:app:eq_1} defines the \emph{gradient} of the energy dependent beta function $\beta(\delta)$, normalized to the unperturbed beta function, i.e. the value of the beta function at $\delta=0$. In order to obtain the actual \emph{change} (beating) of the beta function, in units of the unperturbed beta function, one needs to use the approximation in~\eqref{sec:app:eq0} as:

\begin{equation}
  \frac{\beta(\delta) - \beta_0}{\beta_0} = \frac{\Delta\beta}{\beta}\,\delta\,\,,
\end{equation}
where $\frac{\Delta\beta}{\beta}$ is defined in~\eqref{sec:app:eq_1}. For the general application of optics measurements, it is sufficient to measure the chromatic gradient of the beta function (chromatic beta-beating), normalized to the unperturbed beta function, and not the actual energy dependent shift of the beta function.

\subsection{Second order chromaticity}

Excluding second order dispersion effects, the chromatic beta-beating $\frac{\Delta\beta_z}{\beta_z}$ and the second order chromaticity $Q''_z$ are coupled together through the relationship~\cite{Luo:2010zzg}:

\begin{equation}
\label{exp:eq:chroma_dbb}
Q''_{x,y} = -\frac{1}{2}Q'_{x,y} + \frac{1}{4\pi}\oint[K_{x,y}\mp D_xS_{x,y}]\Delta\beta_{x,y}~ds\,\,,
\end{equation}
where $\Delta\beta_{x,y}=\frac{\partial\beta_{x,y}}{\partial\delta}$ is the chromatic dependence of the horizontal and vertical beta functions, $K_{x,y}$, $S_{x,y}$ are the magnetic strengths of the quadrupoles and sextupoles respectively, and $D_x$ is the horizontal dispersion along the ring. Assuming that the magnetic strengths are static, one concludes that an increase of the horizontal beta-beating translates in a increase or decrease of the second order chromaticity, according to the plane of reference.

\subsection{Approximating shifts of sextupolar components}

As it has been discussed already, the operation of a superconducting wiggler can alter the optics of a circular accelerator. As a result, the optics parameters will depend on the magnetic of the wiggler $B_w$. In order to estimate the additional sextupolar field that the wiggler superimposes during operation, optics perturbations can be introduced either in the analytical form of linear chromaticity, Eq.~\eqref{eq_chroma}, or of the second order chromaticity, Eq.~\eqref{exp:eq:chroma_dbb}, and with the help of the respective experimental observations, the shifts of the sextupolar components during wiggler operation can be approximated. 

Dropping the $x,y$ indexes, the perturbations of the sextupolar magnetic strength of a circular accelerator $S(s,~B_w)$ and its chromatic beta-beating $\Delta\beta(s,~B_w)$ can be expressed as 

\begin{align}
  \label{exp:eq:pert1}
  S(s, B_w) &= S_0(s) + \delta S(s,B_w) \\
  \label{exp:eq:pert2}
  \Delta\beta(s, B_w) &= \Delta\beta_0(s) + \delta \Delta\beta(s,B_w)\,\,,
\end{align}
where $S_0(s)$ and $\delta S(s, B_w)$ are the nominal sextupolar strength and field dependent perturbation in units of m$^{-3}$, $\Delta\beta_0(s)$ and $\delta \Delta\beta(s,B_w)$ are the nominal chromatic beta-beating, $\delta S(s, B_w)$ is the sextupolar component induced by the wiggler at a field of $B_w$, which is only non-zero in the region of the wiggler, and $\delta \beta(s, B_w)$ is the chromatic beta-beating which is non-zero everywhere in the ring. 

In the following analysis, the effects of beta-beating due to a distribution of quadrupolar errors in the ring, and of the second order dispersion are omitted. 

Plugging in Eq.~\eqref{exp:eq:pert1} in Eq.~\eqref{eq_chroma} one can derive the associated change of linear chromaticity $Q'$. If the length of the wiggler is $L_w$ and if the nominal value of the chromaticity at $B_w=0$~T is $Q'_0$, the value of the field dependent chromaticity can be found from

\begin{align}
\label{app:eq_qbw}
  Q'(B_w) &= Q'_0+\frac{1}{4\pi}\int_0^{L_w}\mp\beta(s) D(s) ~\delta S(s, Bw)~ds \\
  & \approx Q'_0\mp\frac{1}{4\pi}~\beta_w~ D_w ~L_w~\delta S_w(B_w) \\ 
  &\approx Q'_0\mp\frac{1}{4\pi}~\beta_w~ D_w ~\widehat{\delta S_w}(B_w) \,\,,
\end{align}
where $\beta_w$, $D_w$ are the average beta function and dispersion in the wiggler, $\widehat{\delta S_w}$ is the sextupolar integrated along the length of the wiggler in units of m$^{-2}$.

Solving Eq.~\eqref{app:eq_qbw} with respect to the sextupolar perturbation, yields an approximation of the shift of the sextupolar component of a ring due to the operation of the wiggle at $B_w$ field as

\begin{equation}
  \label{app:eq_ds1}
\widehat{\delta S_w}(B_w) \approx \mp~4\pi\frac{\Delta Q'(B_w)}{D_w\beta_w}\,\,,
\end{equation}
where $\Delta Q' (B_w) = Q'(B_w)-Q'_0$, is the shift of the linear chromaticity when the field of the wiggler is $B_w$. The above relationship is similar to the equation used to measure the beta function by changing the magnetic strength of a quadrupole and observing the associated tune-shift~\cite{Minty:2003fz}.

The approximated sextupolar perturbation~$\widehat{\delta S_w}(B_w)$ can be also recovered by introducing both perturbations Eq.~\eqref{exp:eq:pert1} and Eq.~\eqref{exp:eq:pert2} in the expression for the second order chromaticity,~Eq.~\eqref{exp:eq:chroma_dbb}.

 By doing so, the expression of the field dependent second order chromaticity becomes

\begin{align}
\label{app:eq_qqbw} 
   Q''(B_w) &= Q''_0 - \frac{1}{2}\bigl(Q'(B_w)-Q'_0\bigr) \\
  & \mp \frac{1}{4\pi}\int_0^{L_w} \Delta \beta(s,B_w)~D(s)~\delta S (s,B_w)~ds \\ 
  & + \frac{1}{4\pi}\oint \delta \Delta\beta(s,B_w)~[K(s)\mp D(s)~S_0(s)]~ds\,\,,
\end{align}
where the last integral includes only the $B_w$ field dependent shift of the chromatic beta-beating $\delta\Delta \beta(s,B_w)$ and it is performed around the whole ring. In addition, the quadrupolar term $K(s)$ depends on the field of the wiggler via Eq.~\eqref{appB:eqDK}. The integral in the middle of the previous wiggler, which includes the $B_w$ field dependent chromatic beta-beating $ \Delta \beta(s,B_w)$ from Eq.~\eqref{exp:eq:pert2}, is performed along the length of the wiggler where the sextupolar perturbation $\delta S (s,B_w)$ is non-zero.
In the case that the quadrupolar strength $K(s)$ of the ring is not given, it can be assumed that the shift of the chromatic beta-beating $\delta\Delta \beta(s,B_w)$, due to a wiggler operating at $B_w$ field, is small. However, this operation might significantly decrease accuracy in the approximation of the sextupolar perturbation. 

By assuming that the last integral of Eq.\eqref{app:eq_qqbw} is small with respect to other terms, the integrated sextupolar shift $\widehat{\delta S_w}$ is found to be

\begin{equation}
\label{app:eq_ds2}
\widehat{\delta S_w}(B_w) \approx \mp~4\pi\frac{\Delta Q''(B_w)+\frac{1}{2}~\Delta Q'(B_w)}{D_w\beta_w\bigl\langle\Delta\beta/\beta\bigr\rangle_w}\,\,,
\end{equation}
where $\Delta Q''(B_w)=Q(B_w)''- Q''_0$ is the shift of the second order chromaticity due to the wiggler field $B_w$, $\bigl\langle\Delta\beta/\beta\bigr\rangle_w$ is the average chromatic beta-beating inside the region of the wiggler, defined in Eq.~\eqref{sec:app:eq_dbb}.
 
In general, Eq.~\eqref{app:eq_ds2} is less accurate than Eq.~\eqref{app:eq_ds1} due to the assumption of a small shift of the chromatic beta-beating and the presence of more parameters such as the linear chromaticity. Under the same assumptions, another estimation of the average chromatic beta-beating at the region of the wiggler that can be analytically deduced from equating Eq.~\eqref{app:eq_ds1} and Eq.~\eqref{app:eq_ds2} and carrying on the algebra is

\begin{equation}
\label{app:eq_dbb_dq}
\bigl\langle\Delta\beta/\beta\bigr\rangle_w \approx \frac{1}{2} + \frac{\Delta Q ''}{\Delta Q'}\,\,.
\end{equation}

\subsection{Error propagation of the chromaticity and chromatic beta-beating measurements\label{sec:app:error}}

By using the formulas:

\begin{align}
\label{sec:app:eq2}
Q'_{z} &= \pm \frac{Q_{s}}{\sigma_{\delta}}\sqrt{\frac{A_{1}+A_{-1}}{A_{0}}}  \\ 
\label{sec:app:eq3}
\frac{\Delta\beta_z}{\beta_{z}}&=\pm \frac{2\,Q'_{z}}{Q_{s}}\biggl|\frac{A_1-A_{-1}}{A_1+A_{-1}}\biggr|
\end{align}
where z=x,~y the horizontal and vertical planes respectively, $Q_z'$ the chromaticity, $Q_s$ the synchrotron tune, $\sigma_{\delta}$ the RMS energy spread, $A_0$ the amplitude of the main betatron line, and $A_{\pm1}$ the chromatic sidebands that appear around $A_0$, one can estimate the chromaticity $Q'_z$ and chromatic beta-beating $\frac{\Delta\beta_z}{\beta_{z}}$, as it is suggested in this paper. In the following analysis, new symbols are introduced for the chromatic beta-beating and the chromatic ratios as:

\begin{align}
\label{sec:app:eq4}
\frac{\Delta\beta_z}{\beta_{z}}&\equiv\Delta\beta \\
\sqrt{\frac{A_{1}+A_{-1}}{A_{0}}}&\equiv R_0 \\
\biggl|\frac{A_1-A_{-1}}{A_1+A_{-1}}\biggr|&\equiv R_{1}
\end{align}
for reasons of convenience. In addition the index of the chromaticity is dropped and it is thus symbolized as $Q'$, since the expressions are valid for both planes.

The measurement errors of chromaticity $\sigma_{Q_z}$ and chromatic beta-beating $\sigma_{\Delta\beta}$ are simply given by:

\begin{align}
\label{sec:app:eq5}
\sigma_{Q'}^2 &= \biggl(\frac{\partial Q'}{\partial Q_s}~\sigma_{Q_s}\biggr)^2 + \biggl(\frac{\partial Q'}{\partial \sigma_{\delta}}~\sigma_{\sigma_{\delta}}\biggr)^2+\biggl(\frac{\partial Q'}{\partial R_0}~\sigma_{R_0}\biggr)^2 \\ 
\label{sec:app:eq6}
\sigma_{\Delta\beta}^2&=\biggl(\frac{\partial \Delta\beta}{\partial Q_s}~\sigma_{Q_s}\biggr)^2 + \biggl(\frac{\partial \Delta\beta}{\partial Q'}~\sigma_{Q'}\biggr)^2+\biggl(\frac{\partial \Delta\beta}{\partial R_1}~\sigma_{R_1}\biggr)^2
\end{align}
where $\sigma_{Q_s}$ is the error in the synchrotron tune $Q_s$ measurement, $\sigma_{\sigma_{\delta}}$ the error in the RMS energy spread $\sigma_{\delta}$ measurement, $\sigma_{R_0}$ the error of the chromatic ratio $R_0$, and $\sigma_{R_1}$ the error of the chromatic ratio $R_1$.

By computing the partial derivatives that are present in Eq.~\eqref{sec:app:eq5} as:

\begin{align}
\label{sec:app:eq7}
\frac{\partial Q'}{\partial Q_s} &= \frac{R_0}{\sigma_{\delta}} = \frac{Q'}{Q_s} \\
\frac{\partial Q'}{\partial \sigma_{\delta}} &= -\frac{R_0~Q_s}{\sigma_{\delta}^2}=-\frac{Q'}{\sigma_{\delta}} \\
\frac{\partial Q'}{\partial R_0} &= \frac{Q_s}{\sigma_{\delta}} = \frac{Q'}{R_0}
\end{align}
 and continuing with the derivatives in Eq.~\eqref{sec:app:eq6} as:

\begin{align}
\label{sec:app:eq8}
\frac{\partial \Delta\beta}{\partial Q_s} &=-\frac{Q'~R_1}{Q_s^2} = - \frac{\Delta\beta}{Q_s} \\
\frac{\partial \Delta\beta}{\partial Q'} &= \frac{R_1}{Q_s}=\frac{\Delta\beta}{Q'} \\
\frac{\partial \Delta\beta}{\partial R_1} &= \frac{Q'}{Q_s} = \frac{\Delta\beta}{R_1}
\end{align}
results in the necessary expressions to calculate Eq.~\eqref{sec:app:eq5} and Eq.~\eqref{sec:app:eq6}.

However, an intermediate step is the calculation of the errors of the chromatic sidebands $\sigma_{R_0}$ and $\sigma_{R_1}$ which is given by the expressions:

\begin{align}
\label{sec:app:eq9}
\sigma_{R_0}^2 &= \biggl(\frac{\partial R_0}{\partial A_0}\sigma_{A_0}\biggr)^2+\biggl(\frac{\partial R_0}{\partial A_1}\sigma_{A_1}\biggr)^2+\biggl(\frac{\partial R_0}{\partial A_{-1}}\sigma_{A_{-1}}\biggr)^2 \\
\label{sec:app:eq10}
\sigma_{R_1}^2 &= \biggl(\frac{\partial R_1}{\partial A_1}\sigma_{A_1}\biggr)^2+\biggl(\frac{\partial R_1}{\partial A_{-1}}\sigma_{A_{-1}}\biggr)^2
\end{align}
where $\sigma_{A_{0}}$, $\sigma_{A_{1}}$ and $\sigma_{A_{-1}}$ are the measurement errors of the amplitudes $A_0$, $A_1$ and $A_{-1}$ respectively. With no loss of generality, the errors in measuring $A_1$ and $A_{-1}$, which mostly come from the limitations of the BPM system and the noise in the TbT signal, can be considered similar i.e. $\sigma_{A_{-1}}\approx\sigma_{A_{1}}\equiv\sigma_{A_{\pm1}}$. By computing the partial derivatives:

\begin{align}
\label{sec:app:eq11}
\frac{\partial R_0}{A_0} &= - \frac{R_0}{2A_0} \\
\label{sec:app:eq12}
\frac{\partial R_0}{A_1} &= \frac{\partial R_1}{A_{-1}} = \frac{1}{2A_0R_0} \\
\label{sec:app:eq13}
\frac{\partial R_1}{A_1} &= \frac{2A_{-1}}{(A_1+A_{-1})^2} \\
\label{sec:app:eq14}
\frac{\partial R_1}{A_{-1}} &= \frac{2A_{1}}{(A_1+A_{-1})^2}\,\,,
\end{align}
the errors $\sigma_{R_0}$ and $\sigma_{R_1}$ in Eq.~\eqref{sec:app:eq9} and Eq.~\eqref{sec:app:eq10} are calculated as:

\begin{align}
\label{sec:app:eq15}
\sigma_{R_0}^2 &= \frac{R_0^2}{4A_0^2}~\sigma_{A_0}^2+\frac{1}{2A_0^2R_0^2}~\sigma_{A_{\pm1}}^2\\\nonumber\\
\label{sec:app:eq16}
\sigma_{R_1}^2 &= 4\frac{A_{1}^2+A_{-1}^2}{\bigl(A_{1}+A_{-1}\bigr)^4}~\sigma_{A_{\pm1}}^2\\\nonumber\\
&=4\biggl(\frac{\sigma_{A_{\pm1}}}{A_1+A_{-1}}\biggr)^2\biggl[1-\frac{2A_1 A_{-1}}{(A_1+A_{-1})^2}\biggr]\,\,.
\end{align}
Concretely, the \emph{normalized} errors in the chromaticity $\sigma_{Q'}$, and chromatic beta-beating $\sigma_{\Delta\beta}$ measurements are found to be:

\begin{align}
\label{sec:app:eq17}
\biggl(\frac{\sigma_{Q'}}{Q'}\biggr)^2 =&~\biggl(\frac{\sigma_{Q_s}}{Q_s}\biggr)^2 + \biggl(\frac{\sigma_{\sigma_{\delta}}}{\sigma_{\delta}}\biggr)^2 +\nonumber  \\ 
&\frac{1}{4}\biggl(\frac{\sigma_{A_0}}{A_0}\biggr)^2 + \frac{1}{2}\biggl(\frac{\sigma_{A_{\pm1}}}{A_1+A_{-1}}\biggr)^2 \\\nonumber \\ 
\label{sec:app:eq18}
\biggl(\frac{\sigma_{\Delta\beta}}{\Delta\beta}\biggr)^2 =&~\biggl(\frac{\sigma_{Q_s}}{Q_s}\biggr)^2 + \biggl(\frac{\sigma_{Q'}}{Q'}\biggr)^2 +\nonumber \\
&~4\biggl[1-\frac{2A_1A_{-1}}{(A_1+A_{-1})^2}\biggr]\biggl(\frac{\sigma_{A_{\pm1}}}{A_1-A_{-1}}\biggr)^2\,\,.
\end{align}
The quadratic terms in the previous expressions testify that, in the presence of measurement errors of the Fourier amplitudes $A_0$, $A_1$ and $A_{-1}$, the chromaticity error $\sigma_{Q'}$ increases for a vanishing betatron amplitude $A_0$ and a small sum of the synchrotron sidebands $A_1$ and $A_{-1}$. Since the amplitude of the synchrotron sidebands depends on the quantity $s=\frac{Q'\sigma_\delta}{Q_s}$, as it is shown in Sec.~\ref{sec:Anal}, therefore it cannot be altered except by changing the beam dynamics parameters, one could experimentally use a sufficiently large excitation in order to increase the $A_0$ term. However, care has to be taken as to be influenced by non-linearities as less as possible, since they might perplex the results.

Moreover, it is evident that similar in amplitude synchrotron sidebands, $A_1$ and $A_{-1}$, penalize the error $\sigma_{\Delta\beta}$ of the chromatic beta-beating $\Delta\beta$ measurement, since this would mean that the chromatic beta-beating is itself very small. Finally, as it is also found for the chromaticity measurement error, the error $\sigma_{\Delta\beta}$ of the chromatic beta-beating increases for a small sum of amplitudes $A_1$ and $A_{-1}$.

\subsection{Impact of a wiggler on optics\label{sec:app:wiggler}}
The 3-D Hamiltonian that describes the motion of an on-momentum particle in the field of a wiggler, with sinusoidal field variation, and expanded up to fourth order is~\cite{Gao:2003tj}

\begin{align}
\label{appB:eqHam}
  \mathcal{H}(x,y,p_x,p_y,s)&=\frac{p_x^2+p_y^2}{2}+\frac{1}{4k^2\rho_w^2}(k_x^2x^2+k_y^2y^2)\nonumber \\
  &+\frac{1}{12k^2\rho_w^2}(k_x^4 x^4+k_y^4 y^4+3k^2k_x^2x^2y^2) \nonumber \\
  &-\frac{\sin(ks)}{2k\rho_w}(p_x(k_x^2x^2+k_y^2y^2)-2k_x^2p_yxy)\,\,,
 \end{align} 
where $x$ refers to the horizontal plane, $y$ to the vertical plane, $k_x^2+k_y^2=k^2=(\frac{2\pi}{\lambda_w})^2$ is the wave-number of the wiggler with $\lambda_w$ the wiggler period, $\rho_w=\frac{B_0\rho_0}{B_w}$ the bending radius of the wiggler with peak field $B_w$ in a ring of magnetic rigidity $B_0\rho_0$, and $s$ is is the parameter that describes the azimuthal motion of the beam. Note that in the case of ideal wiggler with magnet poles of infinite length, $k_x\rightarrow0$ and the focusing of the particle is entirely on the vertical plane.

Inspection of Hamiltonian $\mathcal{H}$, leads to the conclusion that a particle encounters focusing forces inside the wiggler which scale with $\rho_w^{-2}\propto B_w^2$. Due to non-linearities, which arise from the geometry of the wiggler and the initial conditions of the beam, feed-down of the octupoles can generate sextupolar fields, which can have an effect on the beam dynamics.

In the linear regime, the most important contributions come from the quadrupolar terms of Eq.~\eqref{appB:eqHam}, where after differentiation and averaging over the whole length of the wiggler, the additional quadrupolar magnetic strength $\Delta K_z$ per unit cell of the wiggler is

\begin{equation}
  \label{appB:eqDK}
  \Delta K_z\approx-\frac{1}{2}\biggl(\frac{k_z}{kB_0\rho_0}\biggr)^2B_w^2\,\,,
\end{equation}
with $z=x,y$ the horizontal and vertical plane respectively. 

\subsubsection{Tune-shift\label{sec:app:tshift}}

The theoretical tune-shift $\Delta Q_z(B_w)$ due to the presence of the wiggler field $B_w$, which induces the extra quadrupolar component in Eq.~\eqref{appB:eqDK}, is

\begin{align}
\label{appB:eq1}
\Delta Q_z(B_w)&=\frac{1}{4\pi}\int^{L_w/2}_{-L_w/2}\beta_z(s)\Delta K_z~ds\nonumber \\
&=\frac{L_w\langle\beta_z\rangle}{8\pi}\biggl(\frac{k_z}{kB_0\rho_0}\biggr)^2B^2_w\,\,,
\end{align}
where $z=x,~y$ the horizontal or vertical transverse planes. For an ideal wiggler, the focusing is purely vertical ($z=y$) since $k_x=0$, and due to the relationship $k_y^2=k^2$, Eq.~\eqref{appB:eq1} becomes

\begin{equation}
\label{appB:eq2}
\Delta Q_y(B_w)=\frac{L_w\langle\beta_y\rangle}{8\pi B_0^2\rho_0^2}B_w^2
\end{equation}

\subsubsection{Beta-beating}

The presence of a wiggler in a circular accelerator induces perturbation to the linear optics in the form of beta-beating due to the wiggler generated quadrupolar error, Eq.~\eqref{appB:eqDK}. In theory, only vertical perturbations are allowed in the ideal case. The theoretical maximum vertical beta-beating due to the presence of the wiggler is~\cite{katoh1987effect}

\begin{equation}
\label{appB:eq3}
\biggr( \frac{\Delta\beta_y}{\beta_y} \biggl)_{max}=\frac{2\pi\Delta Q_y}{\sin(2\pi Q_y)}\,\,,
\end{equation}
where $Q_y$ is the unperturbed betatron tune, and $\Delta Q_y$ is the tune shift due to the finite wiggler field, given in Eq.~\eqref{appB:eq2}. 

\section{Optics measurements at KARA\label{sec:app:opt}}

In the following sections, the results from optics measurements at the KARA ring, under the influence of the CLIC SC wiggler are presented in order to establish quantitatively and qualitatively the impact of the wiggler on the linear beam dynamics at KARA. The agreement between experiment and theory is also examined, in order to validate the efficiency of the CLIC SC wiggler models that are currently used.

\subsection{Tune-shift due to the CLIC Wiggler\label{app:sec1}}

\begin{figure}[!htb]
  \centering
  \includegraphics[width=0.55\textwidth]{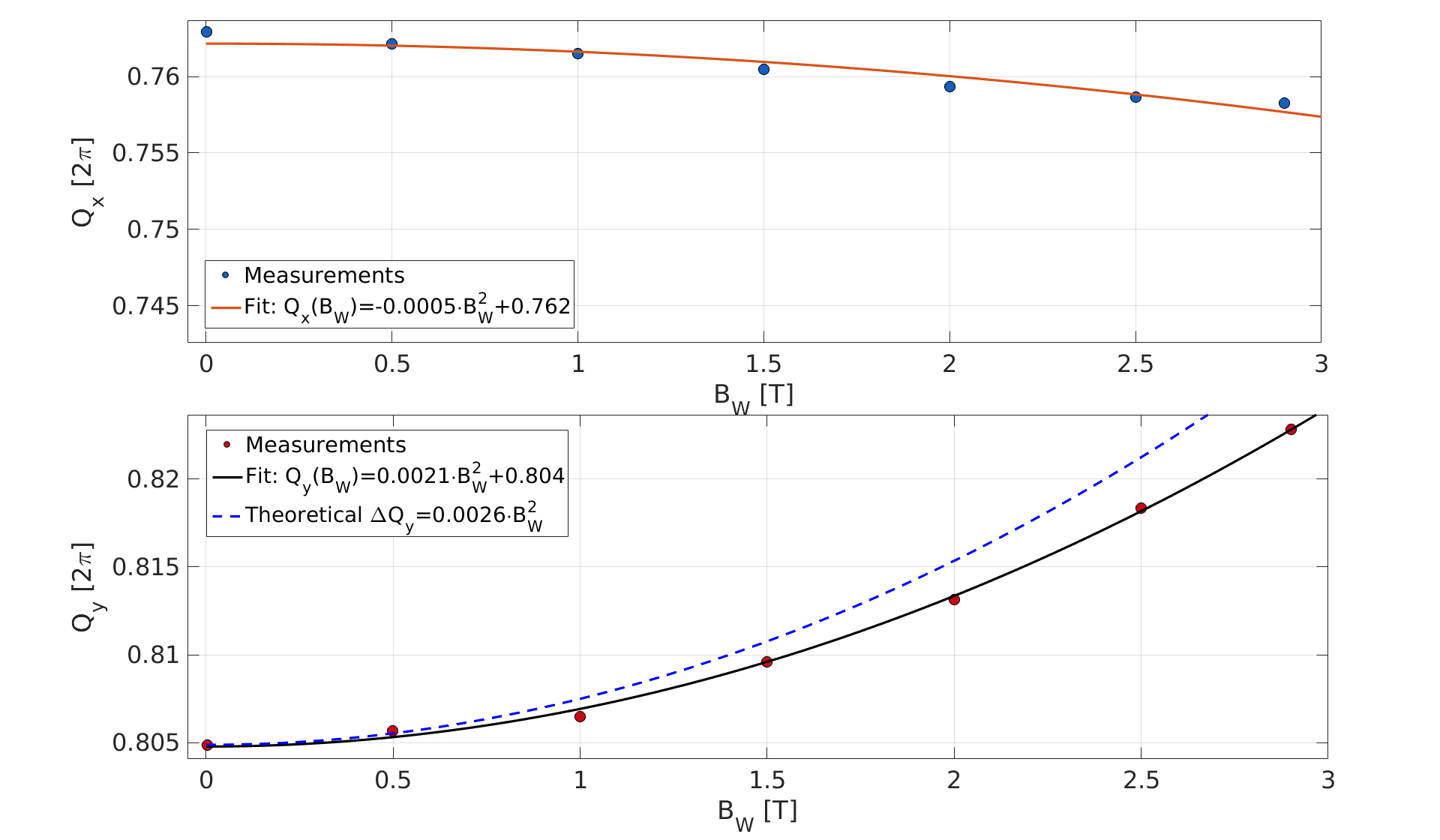}
  \caption{Betatron tune-shift in the horizontal (top) and vertical (bottom) planes at KARA, as a function of the magnetic field of the CLIC SC wiggler. The fits are performed with quadratic models. For the vertical plane, the theoretical tune-shift is shown in black.}
  \label{exp:FIG22}
\end{figure}

 For the KARA ring and the CLIC SC wiggler, the expected vertical betatron tune-shift $\Delta Q_y(B_w)$, by virtue of Eq.~\eqref{appB:eq3}is calculated to be 

\begin{equation}
\label{appB:eq4}
  \Delta Q_y(B_w)=2.6\cdot10^{-3}B^2_w\,\,,
\end{equation}
which is indeed small due to the small vertical beta function at the position of the CLIC SC wiggler. In order to experimentally measure the betatron tune-shift, tune measurements are performed with the mixed BPM method~\cite{PhysRevAccelBeams.22.071002} for around $N=50$ turns. The results are presented in Fig.~\ref{exp:FIG22}, for the horizontal (top) and vertical (bottom) tunes with respect to the increasing magnetic field $B_w$ of the CLIC SC wiggler.

For the evolution of the horizontal tune, a slight shift is observed, with a magnitude that is estimated with a fit of the measured tunes (blue curve) to Eq.~\eqref{appB:eq1}, (orange curve). Although the observed tune-shift, is not dangerous for the operation and beam quality at KARA, it proves the existence of non-linear fields at the location of the CLIC SC wiggler.

Concerning the vertical tune-shift, a similar fit is performed (black curve) in order to estimate its magnitude, which is found similar to the theoretical expectations (blue line), quoted in Eq.~\eqref{appB:eq4}. 

The tune measurements reveal a total normalised tune-shift of around $\Delta Q_x/Q_x=0.5\%$ for the horizontal plane, and roughly $\Delta Q_y/Q_y=2\%$, for the wiggler at maximum field i.e. $B_w=2.9$~T.

\subsection{\label{sec:exp:secsec:betabeat}Quadrupolar beta-beating due to the CLIC Wiggler\label{app:sec2}}

\begin{figure}[!htb]
\centering
\subfloat[Horizontal beta-beating induced by the CLIC SC wiggler at maximum field.\label{exp:FIG34}]{\includegraphics[width=\mysize\textwidth]{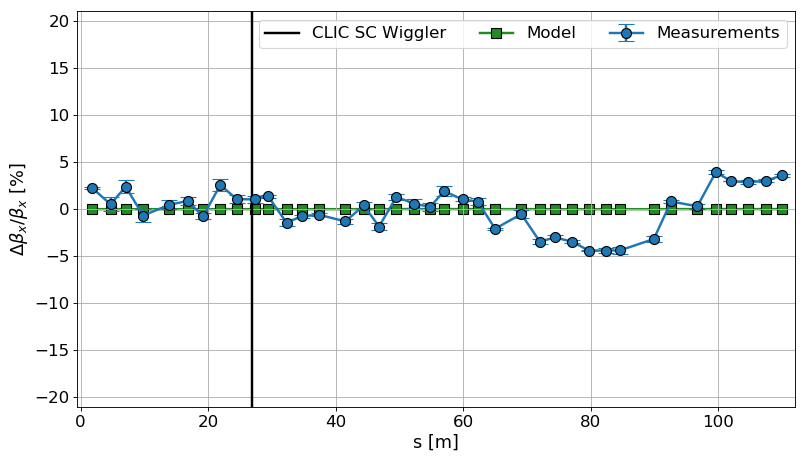}} \\
\subfloat[Vertical beta-beating induced by the CLIC SC wiggler at maximum field. \label{exp:FIG35}]{\includegraphics[width=\mysize\textwidth]{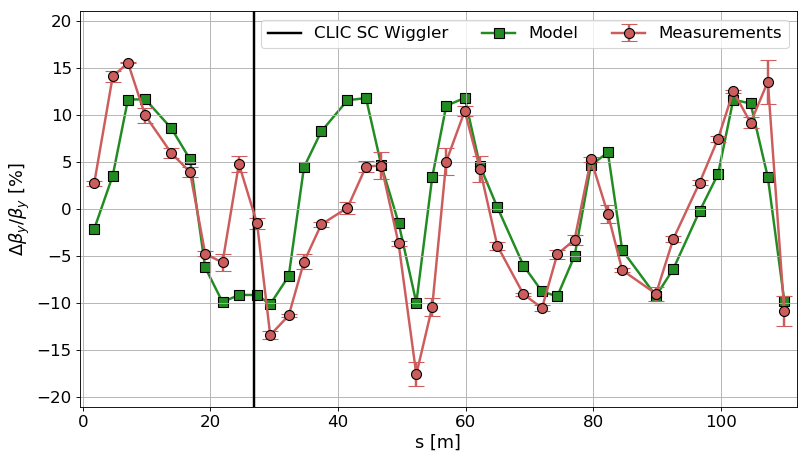}} 

\caption{Comparison of the experimental and model beta-beating, induced by the maximum of the CLIC SC wiggler, with respect to the azimuthal position of the BPMs. The horizontal beta-beating is shown in (a), for the model (green) and experimental measurements (blue). The results for the vertical plane are shown in (b), with the model estimation shown in green, and the experimental measurements in red. The azimuthal position of the CLIC SC wiggler is marked with a black line.}
\end{figure}

For the current experimental measurements with the operation of the CLIC SC wiggler, TbT data are recorded at the $M=39$ BPMs, while ramping up the CLIC SC wiggler in a series of steps, from $0~T$ to $2.9~T$. The estimations of the beta-beating are performed by using information on the Fourier amplitudes of the oscillations~\cite{Zisopoulos:2013esa}, measured with \emph{PyNAFF}, for no wiggler field ($B_w=0$~T) and maximum wiggler field ($B_w=2.9$~T). 

In order to estimate the beta-beating produced by errors of the KARA quadrupoles at the maximum wiggler field, the CLIC SC wiggler is inserted in the KARA model in \emph{ELEGANT}~\cite{Borland:2000gvh} tracking code, and the response of the beta function is recorded for both values of the field. The beta-beating of the model due to the maximum field of the wiggler from the model can be calculated as

\begin{equation}
\label{appB:eq5}
  \frac{\Delta\beta}{\beta}^{model}=\frac{\beta_{W}^m-\beta_0^m}{\beta_0^m}\biggl|_{W=2.9~T}
  \,\,,
\end{equation}
where $\beta_0^m$ is the value of the model beta-function for $B_w=0$~T, and $\beta_W^m$ is the value of the model beta-function for $B_w=2.9$~T.

In order to compare with the experimental estimation of the beta-beating generated purely by the wiggler, one needs first to disentangle the baseline beta-beating, which originates from just the quadrupolar errors in the lattice. The amplitudes of the beam for both cases of $B_w=0$~T and $B_w=2.9$~T are fitted with the model's beta functions at $B_w=0$~T. From this operation, two sets of model dependent, but experimentally measured, beta functions become available, at both fields. Then the experimental beta-beating is found as the difference of the two sets of beta functions, in a manner similar to Eq.~\eqref{appB:eq5}. The difference cancels out the term of the beta-beating due to the quadrupolar errors of the lattice.

The previous considerations can be visualized in Fig.~\ref{exp:FIG34} for the horizontal plane, and in Fig.~\ref{exp:FIG35} for the vertical plane, where the estimated wiggler dependent beta-beating is plotted with respect to the azimuthal position of the BPMs. In both plots, the position of the CLIC SC wiggler is marked with a black line.

The model horizontal beta-beating (green curve) is vanishing as expected, with the experimental measurements (blue curve) agreeing well. A portion of the measurements above $s=60$~m, exhibit an irregular pattern of about $5\%$ RMS value, which is likely coming from calibration issues of these particular BPMs.

Concerning the vertical beta-beating, the agreement between the model (green curve) and the experimental measurements (red curve) is very good. Some deviations are present but they are attributed to the calibration of the BPMs. The experimentally measured vertical beta-beating is less than $15~\%$ in agreement to the theoretical predictions of Eq.~\eqref{appB:eq3}. As expected, due to the small value of the vertical beta-function at the position of the CLIC SC wiggler constraints the vertical beta-beating in relatively low values, which are of no concern for beam stability at KARA.

\bibliographystyle{unsrt}  
\bibliography{Chroma_arxiv.bib}

\end{document}